\documentclass[aps,showpacs,groupedaddress]{revtex4}
\usepackage{amsmath,amssymb,graphicx,float}
\usepackage{dsfont}
\usepackage{soul}
\usepackage{color}
\usepackage{slashed}
\usepackage[utf8]{inputenc}

\usepackage{ulem}
\usepackage{hyperref}

\newcommand*\wbar[1]{\overline{#1}}

\newcommand{\rmRe}{\mathrm{Re}}
\newcommand{\rmIm}{\mathrm{Im}}

\newcommand{\beq}{\begin{equation}}
\newcommand{\eeq}{\end{equation}}
\newcommand{\eexp}{\mathrm{e}}
\newcommand{\I}{\mathrm{i}}
\newcommand{\Tr}{\mathrm{Tr}}

\newcommand{\QCDI}{QCD$_{I}$}
\newcommand{\QCDB}{QCD$_{B}$}
\newcommand{\QCDIl}{QCD$_{I,\lambda}$}
\newcommand{\QCDBl}{QCD$_{B,\lambda}$}

\begin{document}

\title{Applying constrained simulations for low temperature lattice QCD at finite baryon chemical potential}
\author{G. Endr\H{o}di$^{1}$}
\author{Z. Fodor$^{2,3,4}$}
\author{S.D. Katz$^{4,5}$}
\author{D. Sexty$^{2,3,4}$}
\author{K.K. Szab\'o$^{2,3}$}
\author{Cs. T\"or\"ok$^{3,4,5}$}

\affiliation{ $^1$ {\it Institute for Theoretical Physics, Goethe University of Frankfurt, 
Max-von-Laue-Str. 1, 60438 Frankfurt am Main} }

\affiliation{ $^2$ {\it Department of Physics, Wuppertal University,
Gaussstrasse 20, D-42119 Wuppertal, Germany} }

\affiliation{ $^3$ {\it IAS/JSC, Forschungszentrum J\"ulich, D-52425 J\"ulich, Germany} }

\affiliation{ $^4$ {\it Institute for Theoretical Physics, E\"otv\"os University, 
P\'azm\'any Peter s\'etany 1/A, H-1117 Budapest, Hungary} }

\affiliation{ $^5$ {\it MTA-ELTE Lend\"ulet Lattice Gauge Theory Research Group, Budapest, Hungary} }

\begin{abstract}
We study the density of states method as well as reweighting to explore the low temperature phase diagram of QCD at finite baryon chemical potential.
We use four flavors of staggered quarks, a tree-level Symanzik improved gauge action and four stout smearing steps on lattices with $N_s=4,6,8$ and $N_t=6 - 16$.
We compare our results to that of the phase quenched ensemble and also determine the pion and nucleon masses.
In the density of states approach we applied pion condensate or gauge action density fixing.
We found that the density of states method performs similarly to reweighting.
At $T \approx 100$ MeV, we found an indication of the onset of the quark number density at around $\mu/m_N \sim 0.16 - 0.18$ on $6^4$ lattices at $\beta=2.9$.
\end{abstract}

\maketitle

\section{Introduction}
\label{intro_sec}

Understanding the phase diagram of quantum chromodynamics (QCD) is important for high energy physics, nuclear physics and astrophysics as well.
Lattice QCD provides reliable information about QCD in nonperturbative regions and was used e.g. to determine the nature of the chiral phase transition at zero density, which was found to be an analytic crossover \cite{Aoki:2006we}.
Applying lattice QCD to finite baryon density, however, is hindered by the so called sign problem.
The introduction of nonzero baryon chemical potential makes the Boltzmann factor appearing in the path integral complex, and thus the standard Monte-Carlo algorithm based on importance sampling cannot be applied.
To circumvent this problem, several methods have been devised.
However, these methods seem to have a limited range of applicability in their present status.
For a review, see References \cite{deForcrand:2010ys, Wolff:2010zu, Aarts:2013lcm, Gattringer:2014nxa, Sexty:2014dxa, Borsanyi:2015axp, Ding:2017giu, Langfeld:2016kty}.

In this paper we study the density of states method (DoS), which was proposed in Ref. \cite{Bhanot:1986hi} for the three-dimensional Ising model.
Briefly, the method advises the calculation of the histogram of the energy by using constrained simulations.
Using the results of these constrained simulations one can then build the histogram, and use it to calculate the expectation values of other observables as well. We will discuss the method in detail in Sec. \ref{dos_sec}.
Soon after its proposal, it was applied for models with complex actions \cite{Bhanot:1987nv}, and also to finite density QCD at infinite gauge coupling in Ref. \cite{Gocksch:1988iz}.
Later, the method was used to gain information on the phase structure of QCD at finite gauge couplings.
Ref. \cite{Luo:2001id} studied the flavor number dependence of the results with this method at zero chemical potential.
In Ref. \cite{Fodor:2007vv}, the method was used in 4 flavor staggered QCD mainly at $4^4, 6^4$ lattices at various quark masses, using the Wilson(/plaquette) action, while Ref. \cite{Ejiri:2007ga} used the method for 2 flavors of p4-improved staggered quarks at $16^3 \times 4$ lattices.
However, both works used certain approximations.

Besides studying QCD, the method was used and thorougly investigated in other models like compact QED \cite{Azcoiti:1990ng}, random matrix models \cite{Anagnostopoulos:2001yb, Ambjorn:2002pz, Ambjorn:2004jk, Aoki:2014mta}, Z(N) spin models \cite{Bhanot:1986ku, Gattringer:2015lra, Langfeld:2014nta}, SU(2) and U(1) models \cite{Langfeld:2012ah}, two-color heavy-quark QCD \cite{Langfeld:2013xbf}, SU(3) model with static color charges \cite{Giuliani:2017fss}, etc. In recent years, new improvements, like the LLR algorithm \cite{Langfeld:2012ah} and the functional fit approach (FFA) \cite{Gattringer:2015lra} have been developed. The former uses an iterative procedure, while the latter a sequence of one-parameter fits to reduce the statistical error of the histogram.

In our present work, we employ the density of states method (DoS) for QCD with $N_f=4$ flavors of staggered quarks.
Besides applying it in its standard form without any approximations, we investigate also the possibility of improving the results based on insights from the expected low temperature behavior of QCD.

It has been known for a long time (see e.g. \cite{Kogut:1983ia}), that in zero temperature QCD at finite baryon chemical potential (\QCDB), the quark number density should be zero until the quark chemical potential reaches ($\mu$) the first excitation energy, $M_N/3-E_B$.
Here, $M_N$ is the mass of the lightest baryon, the nucleon, and $E_B$ is the binding energy of nuclear matter.
Early attempts to reproduce this phenomenological expectation have revealed that the onset is at a lower chemical potential, around $m_\pi/2$ \cite{Barbour:1986jf}.
It is also believed that the sign problem becomes severe if $\mu$ goes beyond $m_\pi/2$.
In order to clarify this conflict, several attempts were made \cite{Davies:1990qk, Kogut:1994eq, Lombardo:1995dd, Barbour:1997ej, Barbour:1997bh, Nagata:2012tc, Nagata:2012ad}.
These works used either reweighting from $\mu=0$, or the quenched or the phase quenched approximations.
However the phase quenched theory is equivalent to QCD at finite isospin chemical potential (\QCDI) \cite{Stephanov:1996ki, Alford:1998sd}, while the quenched theory is the zero-flavor limit of this.
The onset at $\mu=m_\pi/2$ can then be interpreted as the consequence of pion condensation in \QCDI.
In other words, by doing a phase quenched simulation at $\mu>m_\pi/2$, the
ground state of the phase quenched theory is very different from the ground state of the full theory, since in the former case, pion condensation takes place above $m_\pi/2$, while in the full
theory no pion condensation should occur.
Thus, in this case one generates unimportant configurations when one uses the phase quenched ensemble as the "source" ensemble for reweighting to "target" ensemble QCD at baryon chemical potential.
To overcome this difficulty, we mention a particular idea, that was proposed in random matrix theory (RMT) \cite{Aoki:2014mta}, where the situation is very similar to the case of QCD.
Ref. \cite{Aoki:2014mta} states that by doing constrained simulations and suppressing the pion condensation reduces the overlap problem and makes the sign problem milder in RMT.
We investigate whether similar improvements can cure the problems in QCD, and we focus primarly on the low temperature region of the phase diagram.

The organization of the paper is as follows.
In Section \ref{dos_sec}, we first review the density of states method in general (Sec. \ref{form_sec}), then discuss the method applied for QCD and also give the definitions of the lattice actions and observables we use (Sec. \ref{act&obs_sec} and \ref{constr_op_sec}).
In Section \ref{res_sec}, we present our numerical results regarding the density of states method and compare it to the results of reweighting from the phase quenched theory.
Section \ref{dos:concl_sec} contains the conclusions.

\section{The density of states method}
\label{dos_sec}

\subsection{Formulation of the method}
\label{form_sec}

Suppose we have an arbitrary quantum field theory with quantum fields $\Phi$ and action $S[\Phi]$.
Then in the path integral formalism, the partition function is 
\beq \label{Z_Phi}
Z = \int D\Phi \: \eexp^{-S[\Phi]},
\eeq
where all fields are symbolized with $\Phi$ in this compact notation.
So in the case of QCD, we include gauge and fermion fields in $\Phi$ as well.
Now we can insert a real Gaussian integral in the path integral since it 
changes only the overall normalization of the integral
\beq
Z = \int D\Phi \eexp^{-S[\Phi]} \: N \int dx \: \eexp^{-\frac{\gamma\Omega}{2}(Q[\Phi]-x)^2},
\eeq
where we parametrized the Gaussian with $\gamma$ and $x$, and $\Omega$ is the 4-volume of the system.
$N \propto \sqrt{\gamma \Omega}$ is an irrelevant normalization factor, and the operator $Q$ can be any operator of the theory.
Interchanging the order of integrations, we can write
\beq \label{Z_integration2}
Z = \int dx \: \int D\Phi \: N \: \eexp^{-S[\Phi]-\frac{\gamma\Omega}{2}(Q[\Phi]-x)^2}.
\eeq
Then, if $Q$ is chosen to be the energy and $\gamma \to \infty$, the Gaussian approximates a Dirac-$\delta$, and the second integral gives what is conventionally called in condensed matter physics the density of states. 
The partition function then naturally shows up as the integral of the density of states over all possible values of the energy.
It is also common to call the second integral of Eq. (\ref{Z_integration2}) as the density of states for finite value of $\gamma$ and any operator $Q$.

Naturally, when doing Monte-Carlo simulations, we use a finite value of $\gamma$ and one can think of the exponent of the Gaussian as a potential term added to the action.
This term then constrains the value of $Q$ close to the minimum of that potential, which is likely to be near $x$, when $\gamma$ is large enough.

For our purposes, we need to be even more general to include a reweighting factor in the method to use it for non-real actions. 
We can write $\exp\{-S[\Phi]\} = w[\Phi] \: g[\Phi]$, where we isolate a positive part $g[\Phi]$, that can be used for Monte-Carlo simulations.
Then the definition of the density of states is
\beq \label{rho_def}
\rho(x) = \int D\Phi \: g[\Phi] \: \eexp^{-\frac{\gamma\Omega}{2}(Q[\Phi]-x)^2}.
\eeq
Expectation values can then be written as
\beq \label{dos_expval}
\langle O \rangle = 
{1\over Z} \int D\Phi e^{-S[\Phi]} O[\Phi]  
=\frac{\int \: D\Phi \: O[\Phi] \: w[\Phi] g[\Phi] \: \int dx \: \eexp^{-\frac{\gamma\Omega}{2}(Q[\Phi]-x)^2}}{\int \: D\Phi \: w[\Phi] g[\Phi] \: \int \: dx \: \eexp^{-\frac{\gamma\Omega}{2}(Q[\Phi]-x)^2}} = \frac{\int dx \: \rho(x) \: \langle O w \rangle_{x}}{\int dx \: \rho(x) \: \langle w \rangle_{x}},
\eeq
where the expectation value with the subscript $x$ refers to the expectation value in the constrained ensemble with that specific $x$ value, according to
\beq \label{constr_expval}
\langle A \rangle_x = \frac{1}{\rho(x)} \int \: D\Phi \: A[\Phi] \: g[\Phi] \: \eexp^{-\frac{\gamma\Omega}{2}(Q[\Phi]-x)^2}
\eeq
for an operator $A$.
As one can observe, $\rho(x)$ is the partition function in the constrained ensemble with weight $g[\Phi]$.
As was mentioned above, in the limit $\gamma \to \infty$, $\rho(x)$ measures the histogram of the operator $Q$.
Direct measurement of the histogram would be very challenging as rarely visited bins will have very bad signal to noise ratio (the statistical errors are proportional to $\sqrt{\rho(x)}$).
Using finite $\gamma$ means a smearing of the histogram on the scale $1/\gamma$.
For a large enough value of $\gamma$, $\rho(x)$ will have a maximum (or several maxima) 
around the expectation value of $Q$ without the fixing term, and quickly decays around that.
We can however measure $\rho(x)$ also through its logarithmic derivative
\beq \label{deriv_ln_rho}
\frac{\partial}{\partial x} \ln \rho(x) = \langle \gamma \Omega (Q - x) \rangle_x
\eeq
by carrying out simulations at various $x$ values.
Using this method we get $\rho(x)$ with exponentially reduced errors as compared to the direct measurement of the histogram.

\subsection{Lattice actions and observables}
\label{act&obs_sec}

The system we are interested in is QCD at finite chemical potential using $N_f$ flavors of staggered fermions, defined by the partition function
\beq \label{Z_B}
Z_B = \int DU \eexp^{-S_g[U]} \det ( \slashed{D}(\mu) + m )^{N_f/4},
\eeq
where $\mu$ is the one-third of the baryon chemical potential $\mu_B$, $S_g[U]$ is the tree-level Symanzik improved gauge action using four smearing steps with $\rho=0.125$.
(For simplicity, the lattice spacing $a$ was set to $1$ in the notations of this and the next two sections.
The subscript in Eq. (\ref{Z_B}) refers to the ensemble in which the partition function or the expectation value is calculated, i.e. here $B$ refers to the fact that a non-zero $\mu_B$ is used.)
The gauge observables we are interested in are the gauge action, the spatial and temporal plaquette averages, and the spatial average of the Polyakov loop
\beq
\langle P \rangle_B = \frac{1}{N_s^3} \left\langle \sum_{{\bf n}} L({\bf n}) \right\rangle_B = \frac{1}{N_s^3} \left\langle \sum_{{\bf n}} \textrm{Tr} \prod_{n_4=0}^{N_t-1} U_4({\bf n}, n_4) \right\rangle_B.
\eeq
The Dirac matrix $M(\mu)= \slashed{D}(\mu) + m $ satisfies the 
$\gamma_5$-hermiticity:
\beq
 M(-\mu) = \gamma_5 M(\mu)^+ \gamma_5, 
\label{gamma_5-herm}
\eeq 
 where for the staggered operator the $\gamma_5$ matrix is represented by $\eta_5= (-1)^{n_x+n_y+n_z+n_t}$, where the $n_i$s are the lattice site indices.
From among the fermionic observables, we study the quark number density and the chiral condensate density, defined as
\beq
\langle n \rangle_B = \frac{T}{V} \frac{\partial \ln Z_B}{\partial \mu} = \frac{T}{V} \frac{N_f}{4} \left\langle \frac{\partial \ln \det M(\mu)}{\partial \mu} \right\rangle_B, \qquad \: \langle \bar{\psi}{\psi} \rangle_B = \frac{T}{V} \frac{\partial \ln Z_B}{\partial m},
\eeq
respectively.
The quark number density needs no renormalization, while the chiral condensate should be renormalized, both multiplicatively and additively.
However, in the present paper we do not carry out the continuum limit, thus we do not need to carry out the renormalization.

As it was mentioned earlier, we cannot directly simulate the theory defined by Eq. (\ref{Z_B}) with the Hybrid Monte Carlo algorithm (HMC).
Thus, for generating configurations we need to change either the algorithm or the integration measure, and in this latter case use reweighting in the DoS formulation.
We proceed with this latter option and choose the phase quenched ensemble for generating configurations.
The phase quenched partition function can be written as
\begin{align} \label{Z_pq}
Z_{I} (\lambda = 0 ) &\equiv \int \hspace{-0.1cm}DU \eexp^{-S_g[U]} | \det (\slashed{D}(\mu)+m) |^{N_f/4} \nonumber \\
&= \int \hspace{-0.1cm}DU \eexp^{-S_g[U]}  \det \big((\slashed{D}(\mu)+m)^\dagger (\slashed{D}(\mu)+m)\big) ^{N_f/8} \nonumber \\
&= \int \hspace{-0.1cm}DU \eexp^{-S_g[U]} \det (\slashed{D}(+\mu)+m)^{N_f/8} \det (\slashed{D}(-\mu)+m)^{N_f/8},
\end{align}
where, for the last equality, $\gamma_5$-hermiticity (Eq. (\ref{gamma_5-herm})) was used.
According to the last line of Eq. (\ref{Z_pq}), the phase quenched theory is equivalent to the theory with $N_f/2$ flavors having $+\mu$ and $N_f/2$ flavors having $-\mu$ chemical potentials, i.e. to QCD at isospin chemical potential.
However, the above integration measure is not \textsl{strictly} positive, it can be zero as well.
\footnote{
Regarding numerics, practically, it would never be zero.
When the Dirac operator has small eigenvalues the matrix inversions are slowed down during the simulation.
}

Therefore, in Monte-Carlo simulations, we use the following partition function \cite{Kogut:2002tm}
\beq \label{Z_I,l}
Z_{I}(\lambda)=\hspace{-0.1cm}\int \hspace{-0.1cm}DU \eexp^{-S_g[U]} \det (\slashed{D}(\tau_3\mu)+m+i\lambda\eta_5\tau_2)^{N_f/8}=\hspace{-0.1cm}\int \hspace{-0.1cm}DU \eexp^{-S_g[U]} \det (M^{\dagger}(\mu)M(\mu) + \lambda^2)^{N_f/8},
\eeq
where we have included a ``pion source'' $\lambda$, which renders the determinant strictly positive.
Here, the $\tau_i$ denotes the Pauli matrices acting in flavor space and \mbox{$\eta_5(x)$ is} the "staggered $\gamma_5$" defined earlier.
The flavor off-diagonal term comes from the introduction of the $\lambda \wbar{\psi} \eta_5 \tau_2 \psi$ term in the action before integrating out fermions,
where $\wbar{\psi} \eta_5 \tau_2 \psi$ is proportional to the operator of the pion condensate.
Due to non-zero $\lambda$, this off-diagonal term explicitly breaks the subset of chiral symmetry, that remained after the introduction of the isospin chemical potential.
The expectation values calculated in the above ensemble are denoted as $\langle . \rangle_{I,\lambda}$.
The probability density of Eq. (\ref{Z_I,l}) is used for the HMC simulations both when generating configurations for reweighting to \QCDB\: and in the density of states method as well, completed with a constraining factor in this latter case (see below).

In order to study the effect of explicit isospin symmetry breaking on the theory
and properly define the pion condensate in QCD with baryon chemical potential, we introduce
\beq \label{Z_B,l}
Z_{B}(\lambda) = \hspace{-0.1cm}\int \hspace{-0.1cm}DU \eexp^{-S_g[U]} \det (\slashed{D}(\mu) + m + i\lambda\eta_5\tau_2)^{N_f/8} = \hspace{-0.1cm}\int \hspace{-0.1cm}DU \eexp^{-S_g[U]} \det (M^{\dagger}(-\mu)M(\mu) + \lambda^2)^{N_f/8},
\eeq
which is the partition function of QCD at baryon chemical potential with explicit isospin symmetry breaking, due to finite $\lambda$.
This theory is referred to as \QCDBl.
With the help of Eq. (\ref{Z_I,l}) and (\ref{Z_B,l}) the pion condensate operators of \QCDIl \:and \QCDBl \: are
\begin{align}
\langle \pi \rangle_{I,\lambda} &= \frac{T}{V} \frac{\partial \ln Z_I(\lambda)}{\partial \lambda} = \frac{T}{V} \frac{N_f}{8} \: 2\lambda \: \left\langle \textrm{Tr}\left( M(\mu)^\dagger M(\mu) + \lambda^2 \right)^{-1} \right\rangle_{I,\lambda}, \label{pi_Il} \\
\langle \pi \rangle_{B,\lambda} &= \frac{T}{V} \frac{\partial \ln Z_B(\lambda)}{\partial \lambda} = \frac{T}{V} \frac{N_f}{8} \: 2\lambda \: \left\langle \textrm{Tr}\left( M(-\mu)^\dagger M(\mu) + \lambda^2 \right)^{-1} \right\rangle_{B,\lambda}, \label{pi_Bl}
\end{align}
respectively.
Somewhat surprisingly, these two operators differ from each other. This is just a simple consequence of integrating out fermions.
Similarly, other observables that are derived with the help of the determinants differ from each other in \QCDBl\: and \QCDIl.
Besides the pion condensate, we study here the behavior of the quark number density in \QCDBl, which is defined as %(the real part of)
\beq
\langle n \rangle_{B,\lambda} = \frac{T}{V} \frac{\partial \ln Z_B(\lambda)}{\partial \mu} = \frac{T}{V} \frac{N_f}{8} \: \left\langle \frac{\partial \ln\det (M(-\mu)^\dagger M(\mu) + \lambda^2)}{\partial \mu} \right\rangle_{B,\lambda}.
\eeq

\subsection{Choosing the operator to constrain}
\label{constr_op_sec}

As was discussed in Sec. \ref{dos_sec}, the physics of the system to be studied can give useful hints as to what operators could be useful to include in the fixing term.
In particular, in \QCDB\: the Silver Blaze phenomenon indicates that at low temperatures the vacuum state should persist until the quark chemical potential hits roughly the third of the nucleon mass.
In the phase quenched (or \QCDIl) simulation, however, configurations with large pion condensate will occur when $\mu$ is over half of the pion mass. Their contribution in the observable has to cancel out eventually, and this may require huge statistics. By naive reweighting from \QCDIl, one can not really avoid these configurations.
Fixing the pion condensate to values near zero, however, could help to suppress the occurence of such undesirable configurations.
Moreover, according to the results that will be presented later, there is a non-zero correlation between the pion condensate and the gauge action density (see Fig. \ref{fig:sg-pi,corr}), which poses the idea to test the DoS method with fixing the latter alone as well.
Therefore, in this study, we have applied the DoS formulation with fixing the pion condensate or the gauge action density.
The implementation of the fixing for the gauge action density is straightforward, so we turn to the pion condensate.

As was mentioned, the operators for measuring the pion condensate in \QCDBl \:and \QCDIl \:differ: $\langle \pi \rangle_{I,\lambda}$ is real, while $\langle \pi \rangle_{B,\lambda}$ is complex in general, which makes the latter more complicated to constrain.
Here, we do not elaborate on this question and continue with constraining the pion condensate of \QCDIl.

Usually, traces in lattice QCD are calculated stochastically, according to
\beq \label{tr,noise_vecs}
  \Tr A \approx \frac{1}{N_v} \sum_{i=1}^{N_v} {\eta^{(i)}}^\dagger A \eta^{(i)}, \quad \textrm{with} \qquad \frac{1}{N_v} \sum_{i=1}^{N_v} { \eta_j^{(i)*} \eta_k^{(i)} } \approx \delta_{jk}
\eeq
where $j$ and $k$ label the components of the noise vector $\eta^{(i)}$, and $N_v$ denotes the number of noise vectors.
Applying this formula for the pion condensate would be very expensive, if one would like to use it to reach reasonable precision when calculating the action for the accept/reject steps.
We can overcome this problem with the help of the $N_{pf}$ complex scalar fields (also called pseudofermion fields) that are used in the calculation of the determinant.
The determinant of Eq. (\ref{Z_I,l}) is represented with these fields in the following way
\beq \label{det_psf}
\det(M^\dagger(\mu)M(\mu) + \lambda^2)^{N_f/8} \propto \int \prod_{j=1}^{N_{pf}} D \phi^{\dagger}_j D \phi_j \exp\left\{ -\sum_{j=1}^{N_{pf}} \phi^\dagger_j (M^\dagger(\mu)M(\mu) + \lambda^2)^{-\frac{N_f}{8N_{pf}}} \phi_j \right\}.
\eeq
Usually, the $\phi_j$ scalar fields are refreshed at the beginning of each trajectory in the HMC algorithm, as they appear only quadratically and can thus
be conveniently generated with the above distribution.
 We can use Eq. (\ref{det_psf}) to give another expression for the pion condensate:
\beq \label{pi_phi}
\langle \pi_{\phi} \rangle_{I,\lambda} = \frac{T}{V} \frac{\partial \ln Z_{I,\phi}(\lambda)}{\partial \lambda} = \frac{T}{V} \left\langle \frac{N_f}{8N_{pf}} \: 2\lambda \: \sum_{j=1}^{N_{pf}} \phi_j^\dagger \left( M(\mu)^\dagger M(\mu) + \lambda^2 \right)^{-\frac{N_f+8N_{pf}}{8N_{pf}}} \phi_j \right\rangle_{I,\lambda}.
\eeq

Note that in Eq. (\ref{pi_phi}), $Z_{I,\phi}(\lambda)$ is equivalent to $Z_I(\lambda)$ (Eq. (\ref{Z_I,l})), the only difference is that the determinant is represented with pseudofermion fields in the former case.
We can now include this operartor in the constraining term of Eq.(\ref{Z_integration2}). In this case the pseudofermionic fields no longer 
appear quadratically, thus a refreshment at the beginning of 
each trajectory using the heatbath is no longer possible.
We use them as dynamical fields in the HMC evolution. 

We note that the operators from (\ref{pi_Il}) and (\ref{pi_phi}) do not give explicitly the same result on a given configuration, only when the number of noise vectors and the number of pseudofermions goes to infinity.

\subsection{Reweighting}
\label{rew_sec}

Before presenting the results, we briefly overview here the reweighting formulas which we use in the DoS and for comparison as well.
For calculating the expectation value of an operator $O$ in \QCDB we use the following formulas
\beq \label{rew1}
Z_B=\langle w_B \rangle_{I,\lambda} Z_I(\lambda), \qquad \langle O \rangle_B = \frac{\langle O w_B \rangle_{I,\lambda}}{\langle w_B \rangle_{I,\lambda}}.
\eeq
Here $Z_B$ and $Z_I(\lambda)$ are given by Eq. (\ref{Z_B}) and (\ref{Z_I,l}), respectively, and $w_B$ denotes the weight.
The logarithm of this weight is given by
\beq \label{w_B}
\ln w_B=\frac{N_f}{4} \left( \ln \det M(\mu) - \frac{1}{2} \ln \det ( M(\mu_0)^\dagger M(\mu_0) + \lambda^2 ) \right).
\eeq
$\mu_0$ denotes the chemical potential, where simulations are carried out and $\mu$ is the chemical potential we reweight to.
Since $\det M(\mu)$ is complex, its logarithm is defined only up to an additive $2k\pi i$ term, where $k$ is an integer.
When $N_f \neq 4$, this means that the weight is not defined unambiguously.
One possibility is to choose the appropriate $k$ by demanding the weight to be a continous function of $\mu$ along the positive real axis  \cite{Fodor:2001pe}, we note, however, that the correctness of this procedure and in more general, the rooting procedure with and without $\mu$ is still, to some extent, under investigation \cite{Golterman:2006rw, Kronfeld:2007ek}.
Nevertheless, the above ambiguity does not affect our reweighting in the aformentioned case, because we use $N_f=4$ in this paper.

But besides that, as was mentioned in Sec. {\ref{act&obs_sec}}, we also reweight to \QCDBl \:at finite $\lambda$ (Eq. \ref{Z_B,l}).
We have various considerations for doing this. 
First, we would like to calculate the pion condensate in QCD at finite baryon chemical potential (Eq. \ref{pi_Bl}), which can be nonzero on average at a finite lattice only when one includes the explicit breaking with $\lambda$.
Second, we would like to see whether the effect of the explicit breaking in \QCDBl \:can make any difference when we carry out the $\lambda \to 0$ extrapolation as compared to reweighting directly to $\lambda=0$.
When reweighting to \QCDBl \:at finite $\lambda$, the logarithm of the weight becomes
\beq \label{w_Bl}
\ln w_{B,\lambda} = \frac{N_f}{8} \Big( \ln\det (M(-\mu)^\dagger M(\mu) + \lambda^2) - \ln\det(M(\mu_0)^\dagger M(\mu_0) + \lambda^2) \Big).
\eeq
Thus, in this case, the above mentioned ambiguity holds on, hence we use the continuity of the weights as a function of $\lambda$ and $\mu$ to choose the appropriate Riemann sheet.
This can be done, however, only if one knows the analytical dependence of the determinant on $\mu$ and $\lambda$.
The former is known in the $\lambda=0$ case, due to the so called reduction formula \cite{Fodor:2001pe}.
Regarding the $\lambda$-dependence, we calculated the eigenvalues of $M(-\mu)^\dagger M(\mu)$, and used these with $\lambda^2$ shifted to obtain the determinant.
For determining the appropriate Riemann surface of $\ln\det(M(-\mu)^\dagger M(\mu) + \lambda^2)$, we fixed the imaginary part of $\ln\det(M(-\mu)^\dagger M(\mu))$ comparing it to $2\ln\det M(\mu)$ -- the latter obtained by the reduction formula --, and then used the same $2k\pi i$ term when we calculate with $\lambda$ via
\beq
\ln\det\left(M(-\mu)^\dagger M(\mu) + \lambda^2\right) = \sum_i {\rm{Ln}}\left(\left(\xi_i + \lambda^2\right) \right) + 2 k\pi \I,
\eeq
where the $\xi_i$s are the eigenvalues of $M(-\mu)^\dagger M(\mu)$ and $2 k\pi i = 2 \rmIm \ln\det M(\mu) - \rmIm \sum_i {\rm{Ln}} \, \xi_i$, where ${\rm{Ln}}$ is the logarithm with imaginary part in between $(-\pi, \pi)$.
One can then use similar formulas as in Eq. (\ref{rew1}), but with the weights of Eq. (\ref{w_Bl}), namely
\beq \label{rew2}
Z_B(\lambda)=\langle w_{B,\lambda} \rangle_{I,\lambda} Z_I(\lambda), \qquad \langle O \rangle_{B,\lambda} = \frac{\langle O w_{B,\lambda} \rangle_{I,\lambda}}{\langle w_{B,\lambda} \rangle_{I,\lambda}}.
\eeq
This procedure is quite expensive -- the computational cost is $\mathcal{O}((N_s^3 N_t)^3)$ --, thus we carried it out only on our smallest lattices.

In the DoS formulation, we used the weights of Eq. (\ref{w_B}) or Eq. (\ref{w_Bl}) in the integrals of Eq. (\ref{dos_expval}), when we calculated the expectation value of an observable $O$, $\langle O \rangle_B$ or $\langle O \rangle_{B,\lambda}$, respectively.
Moreover, by setting the weights to $1$ in Eq. (\ref{dos_expval}), one can calculate $\langle O \rangle_{I,\lambda}$, or $\langle O \rangle_{I}$ by using 
\beq \label{w_I}
\ln w_I=\frac{N_f}{4} \left( \ln |\det M(\mu)| - \frac{1}{2} \ln \det ( M(\mu_0)^\dagger M(\mu_0) + \lambda^2 ) \right).
\eeq
$\langle O \rangle_{I}$ denotes the expectation value of the operator $O$ in \QCDI, which is identical to the phase quenched ensemble.
Using a leading order expansion for the weights of Eq. (\ref{w_I}) (cf. Ref. \cite{Brandt:2016zdy}), one can rewrite $\ln w_B$ as
\begin{align} \label{lo-rew1}
  \ln w_B &= \ln\left( \frac{{\det M(\mu)}^{N_f/4}}{|\det M(\mu_0)|^{N_f/4}}\: \frac{|\det M(\mu_0)|^{N_f/4}}{\det (M(\mu_0)^\dagger M(\mu_0)+\lambda^2)^{N_f/8}} \right) \nonumber \\
  &= \frac{N_f}{4} \ln \left( \frac{\det M(\mu)}{\ln|\det M(\mu_0)|} \right) + \frac{N_f}{8} \ln\left( \frac{\det (M(\mu_0)^\dagger M(\mu_0) + \lambda^2 - \lambda_w^2)|_{\lambda_w^2=\lambda^2}}{\det (M(\mu_0)^\dagger M(\mu_0) + \lambda^2)} \right) \nonumber \\
  &= \frac{N_f}{4} \ln \left( \frac{\det M(\mu)}{\ln|\det M(\mu_0)|} \right) - \frac{N_f}{8} \left( \frac{\partial\ln\det(M(\mu_0)^\dagger M(\mu_0) + \lambda^2)}{\partial\lambda} \frac{\lambda}{2} + \mathcal{O}(\lambda^4) \right) \nonumber \\
  &= \frac{N_f}{4} \ln \left( \frac{\det M(\mu)}{\ln|\det M(\mu_0)|} \right) - \frac{\lambda}{2} \frac{V}{T} \pi + \mathcal{O}(\lambda^4),
\end{align}
where $\pi$ is the pion condensate operator in \QCDIl\:(cf. Eq. (\ref{pi_Il})).
In Eq. (\ref{lo-rew1}) we introduced the parameter $\lambda_w$ and performed Taylor-expansion in it.
On one hand, this formula shows -- at least to leading order in $\lambda$ --, that when one reweights from \QCDIl\: to \QCDB, apart from the phase, a large pion condensate suppresses the weight.
On the other hand, the above formula would be practically useful as well, because estimating the weight by measuring the pion condensate is computationally much cheaper, than calculating $\det(M(\mu_0)^\dagger M(\mu_0)+\lambda^2)$.
However, unfortunately, we found that when one reaches the pion condensation region, the formula is no longer reliable and it overestimates the actual weights (c.f. \cite{Brandt:2017oyy}).
Whether the next term in the Taylor expansion improves the behavior or not is left for further study, i.e. we calculate $\det(M(\mu_0)^\dagger M(\mu_0)+\lambda^2)$ directly by Gauss elimination in the following.

We note here, that one can take into account the fixing term with the help of reweighting as well, according to
\beq \label{rew3}
Z_B = \left\langle w_{B} \exp\left\{\frac{\gamma\Omega}{2}(Q-x)^2\right\}\right\rangle_x \rho(x), \qquad \langle O \rangle_B = \frac{\left\langle O w_B \exp\left\{\frac{\gamma\Omega}{2}(Q-x)^2\right\} \right\rangle_x}{ \left\langle w_B \exp\left\{\frac{\gamma\Omega}{2}(Q-x)^2\right\}\right\rangle_x },
\eeq
where $Q$ is the fixed operator, $\Omega$ is the lattice volume and $\gamma$ is a parameter that controls the width of the Gaussian.
$\rho(x)$ is given by Eq. (\ref{rho_def}) applied for QCD, with the $\Phi$ fields corresponding to the link variables and $g$ chosen to be the integrand of $Z_I(\lambda)$ (see Eq. (\ref{Z_I,l})).
The identities of Eq. (\ref{rew3}) are valid for all $x$.
Although the exponential factor in the expectation values seems to be quite problematic - since there is a volume factor in the exponent -, we investigate whether the distribution of $Q-x$ can be narrow enough to compensate the large $\gamma \Omega /2$ factor.
We refer to this approach as \textsl{direct reweighting from the constrained ensemble} in the following.
If the fluctutation of the exponent could be made small and $Z_B$ as well as $\langle O \rangle_B$ would not depend on $x$ (at least, for a wide enough interval), then the method may provide reasonable results without integration in $x$.

Using this formalism (Equations (\ref{rho_def})-(\ref{deriv_ln_rho})), 
we can do simulations based on importance sampling in the constrained ensemble and with the help of those results we can recover the expectation values in the original ensemble, defined by Eq. (\ref{Z_Phi}).

Since the partition sum can be written as $Z_B=\Tr(\exp\{-(H-\mu N)/T\})$ and the Hamiltonian $H$ commutes with the particle number operator $N$, $Z_B$ is a sum of positive numbers $Z_B=\sum_{n,N} \exp\{-(E_n-\mu N)/T\}$.
As a consistency or reliability criteria, we demand the DoS as well as reweighting to provide a positive $Z_B$ within at least 2 standard deviations.
Every observable will inherit the relative error of $Z_B$ and thus if the measured $Z_B$ is not positive within a few standard deviations then even the magnitude of $Z_B$ is unclear and thus the results will be unreliable.

\section{Results}
\label{res_sec}

\subsection{Simulation details}

We performed simulations with $N_f=4$ flavors of staggered fermions, on $4^3 \times (8,12,16), 6^3 \times (6,8,12)$ lattices.
We used the following quark mass and $\beta$ pairs in the simulations: for $N_s=4$, ($ma$, $\beta$)=\{(0.05, 2.9), (0.02, 2.74)\}, and for $N_s=6$, ($ma$, $\beta$)=\{(0.05, 2.9), (0.02, 2.9)\} was set.
We used several $\lambda a$ and $\gamma$ values at some simulation points to see their effect on the results.
In order to determine the lattice spacing using $w_0$ \cite{Borsanyi:2012zs}, the pion and nucleon masses, we used $N_t=24$ and $N_t=32$ lattices.
The results and the simulation parameters for these runs are summarized in Table \ref{ta:zero_T}.
We also checked that the simulation points are in the confinement region.
The Polyakov loop was small in our simulations also at finite $\mu a$.
  We estimated the pseudocritical $\beta_c$ on a  $16^3 \times 6$ lattice by calculating the renormalized chiral susceptibility (by subtracting the chiral susceptibility measured on a  $16^3 \times 32$ lattice and multipling by the square of the quark mass).
At $m_\pi \approx 335$ MeV, we found $\beta_c$ to be around $\sim 3.36$, which corresponds to $T_c \sim 137$ MeV.
For the larger $m_\pi$, using the same lattice sizes, we found $T_c$ to be in the range 130-160 MeV.
Since the lattices are quite coarse, we applied four stout smearing steps 
using $\rho=0.125$  to reduce the finite lattice spacing effects.
Finite volume effects are expected to be moderate on the lattices with quark mass $ma=0.02$ ($m_\pi a N_s \sim 2.3$) and somewhat smaller on lattices at $ma=0.05$ ($m_\pi a N_s \gtrsim 3$).

\begin{table}[H]
    \centering
    \begin{tabular}{|c|c|c|c|c|c|c|c|c|c|}
        \hline
        $N_s$ & $N_t$ & $\beta$ & $ma$ & $\lambda a$ & $a$[fm] & $m_\pi a$ & $m_\pi$[MeV] & $m_N a$ & $m_N$[MeV] \\
        \hline
        4 & 32 & 2.74 & 0.02 & 0.0 & \:0.33565(6)\: & 0.571(1) & 335.7(6) & 1.73(11) & 1017(65) \\
        4 & 32 & 2.74 & 0.02 & \:0.004\: & - & 0.576(3) & - & - & - \\
        4 & 32 & 2.9 & 0.05 & 0.0 & 0.32876(2) & 0.728(1) & 437.0(6) & 2.03(8) & 1218(48) \\
        4 & 32 & 2.9 & 0.05 & 0.01 & - & 0.730(3) & - & - & - \\
        6 & 24 & 2.9 & 0.02 & 0.0 & 0.33048(2) & 0.381(1) & 227.5(7) & 1.66(15) & 1009(90) \\
        6 & 32 & 2.9 & 0.05 & 0.0 & 0.33304(4) & \:0.565(1)\: & 334.8(7) & 1.76(7) & 1055(17) \\
        6 & 32 & 2.9 & 0.20 & 0.0 & 0.34044(3) & \:1.077(1)\: & 624.2(7) & 2.34(3) & 1356(18) \\
        16 & 32 & 2.9 & 0.02 & 0.0 & 0.33080(3) & \:0.359(2)\: & 214.2(3) & - & - \\
        16 & 32 & 2.9 & 0.05 & 0.0 & 0.33329(3) & \:0.555(1)\: & 328.6(6) & - & - \\
        \hline
    \end{tabular}
\caption{
\label{ta:zero_T}
Summary of zero temperature runs to determine the pion and nucleon mass and the lattice spacing.
We note that the results and errors for the nucleon mass are quite indefinite.
At small quark masses, the relative error for the staggered nucleon correlator grows exponentially with time measured in lattice units, therefore using the accumulated statistics 
the effective mass plateau is only faintly recognized.
}
\end{table}

\subsection{Fixing $\pi_\phi$}
\label{fix_pion_cond}

In this section we present the DoS results that were obtained by constraining the pion condensate, $\pi_\phi$ (Eq. \ref{pi_phi}).
We achieve this as it was discussed at the end of Sec. \ref{constr_op_sec} and we used $N_{pf}=1$ pseudofermion fields in most of our simulations.
We note that by using only one pseudofermion field, in the range $x \in [-0.18,0.18]$, where $x$ refers to the value at which one would like to constrain $\pi_\phi$, the simulations have a tendency to get 'stuck' and even break down because of very large HMC forces, if the timestep is too large.
With a sufficenty small timestep, where the acceptance ratio is larger than $\sim 0.9$, no such problems occur.
No such problem was found by using more than one pseudofermion field, but in these cases the simulation is more expensive.

It is important to recall that $\pi_\phi$ does not give the same result as the pion condensate calculated with the help of noise vectors (Eq. (\ref{pi_Il}), (\ref{tr,noise_vecs})), they are equal only in the limit as the number of pseudofermions as well as the number of noise vectors goes to infinity.
$\langle \pi_\phi \rangle_x$ and $\langle \pi \rangle_{I,x}$ are shown in Fig. \ref{picond,npf} using $N_{pf}=1,2,3$ in the constrained simulations.
After carrying out the $x$ integration according to Eq. (\ref{dos_expval}), we get back the pion condensate of the unconstrained simulation in \QCDIl, where the pion condensate is measured with the operator in Eq. (\ref{pi_Il}).

\begin{figure}[H]
\begin{center}
\includegraphics[scale=0.78]{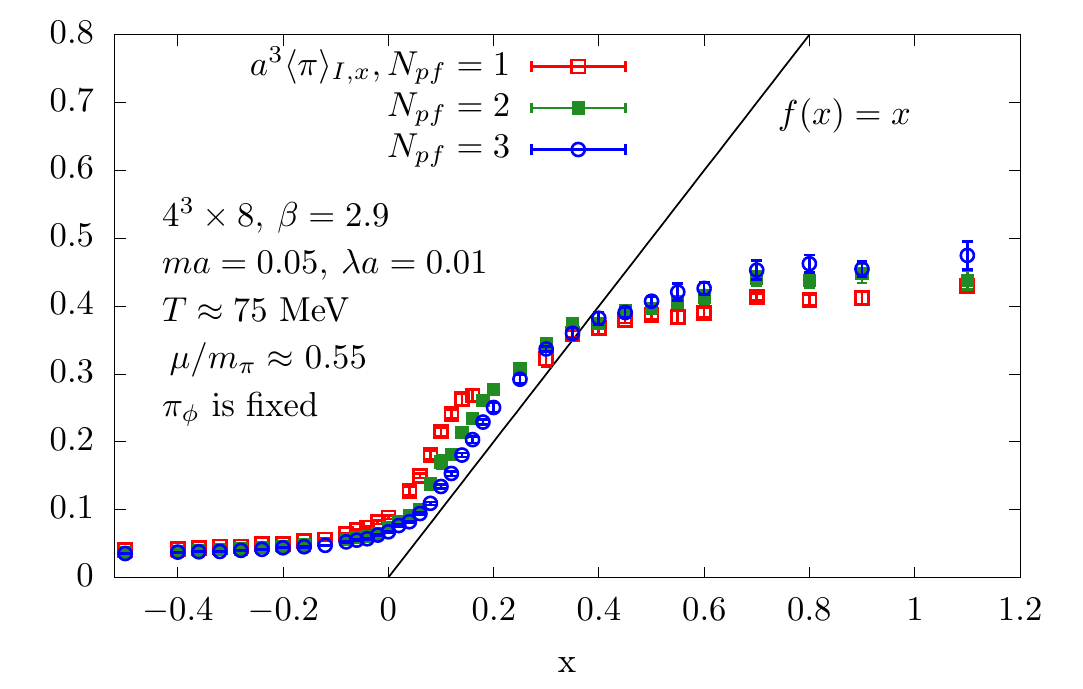}
\includegraphics[scale=0.78]{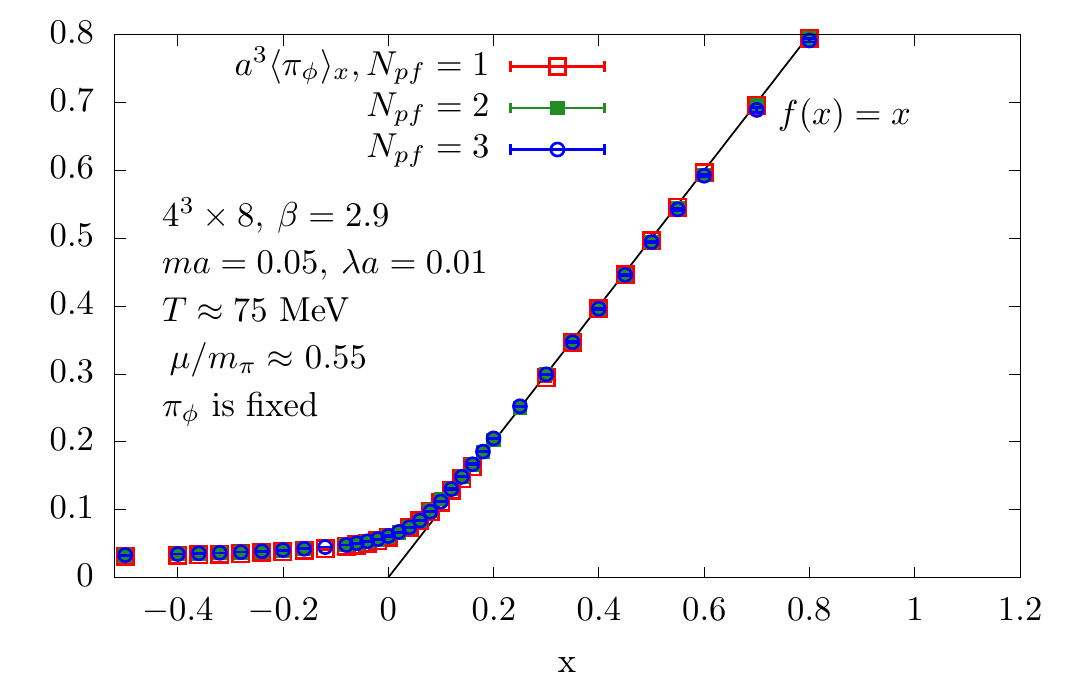}
\caption{The expectation values for the pion condensate in the constrained ensembles using noise vectors ($\langle \pi \rangle_{I,x}$) and with the help of pseudofermions ($\langle \pi_\phi \rangle_x$), left and right panel, respectively.
The simulations were carried out by constraining $\pi_\phi$ using $N_{pf}=1,2,3$.}
\label{picond,npf}
\end{center}
\end{figure}

We emphasize again that both operators for the pion condensate in \QCDIl\: are real and positive definite at finite $\lambda$.
Therefore, one can not constrain $\pi_\phi$ to zero or negative values, but can push it closer to zero e.g. by writing a negative $x$ in the fixing term.
We found that it is more efficient to proceed this way, rather than decreasing the width of the Gaussian of the fixing term when using a small positive $x$ or $x=0$, because a smaller width (larger $\gamma$) results in larger forces and slows down the simulation at other values of $x$ as well.
The DoS setup is validated at $\mu=0$ by fixing $\pi_\phi$ and calculating the full DoS integrals, and ensuring that we obtain results consistent with simulations using no fixing.

\begin{figure}[H]
\begin{center}
\includegraphics[scale=0.85]{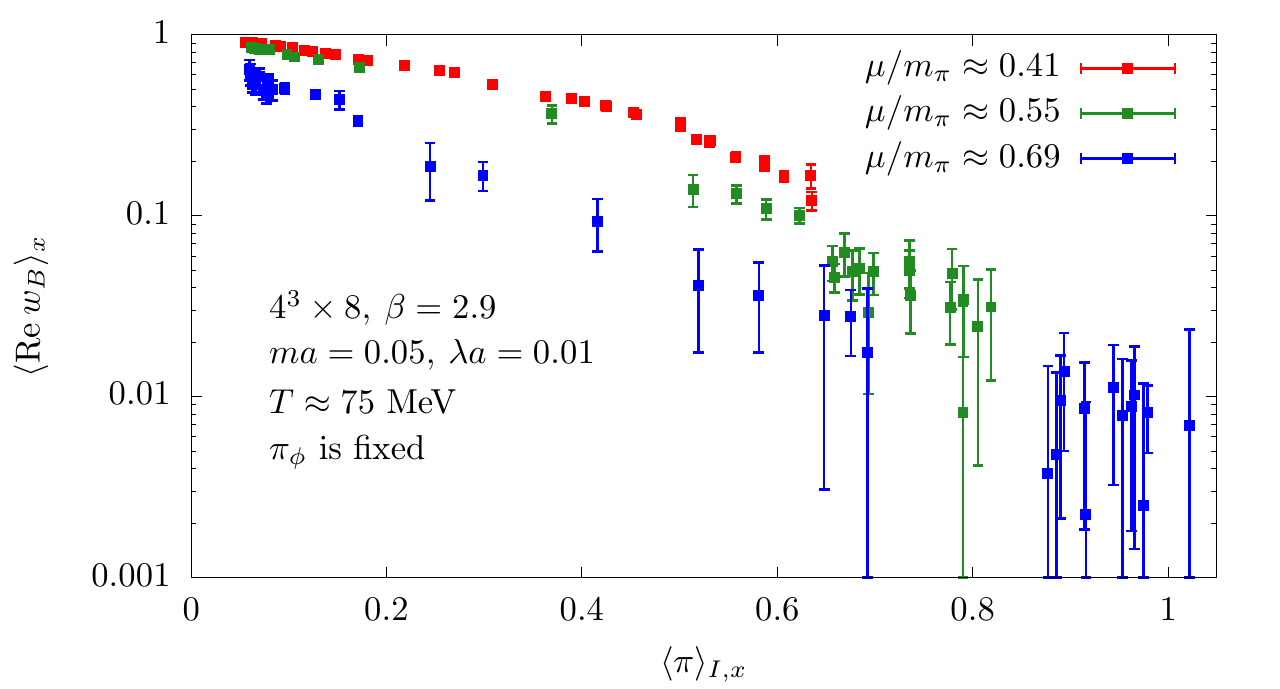}
\caption{
The expectation value of the real part of the weight as a function of the expectation value of the pion condensate $\langle \pi \rangle_{I,x}$, in simulations with fixing $\pi_\phi$ for $\lambda=0.01$ and several $\mu$ values.
}
\label{pic_vs_w_pifix}
\end{center}
\end{figure}

\begin{figure}[H]
\begin{center}
\includegraphics[scale=0.75]{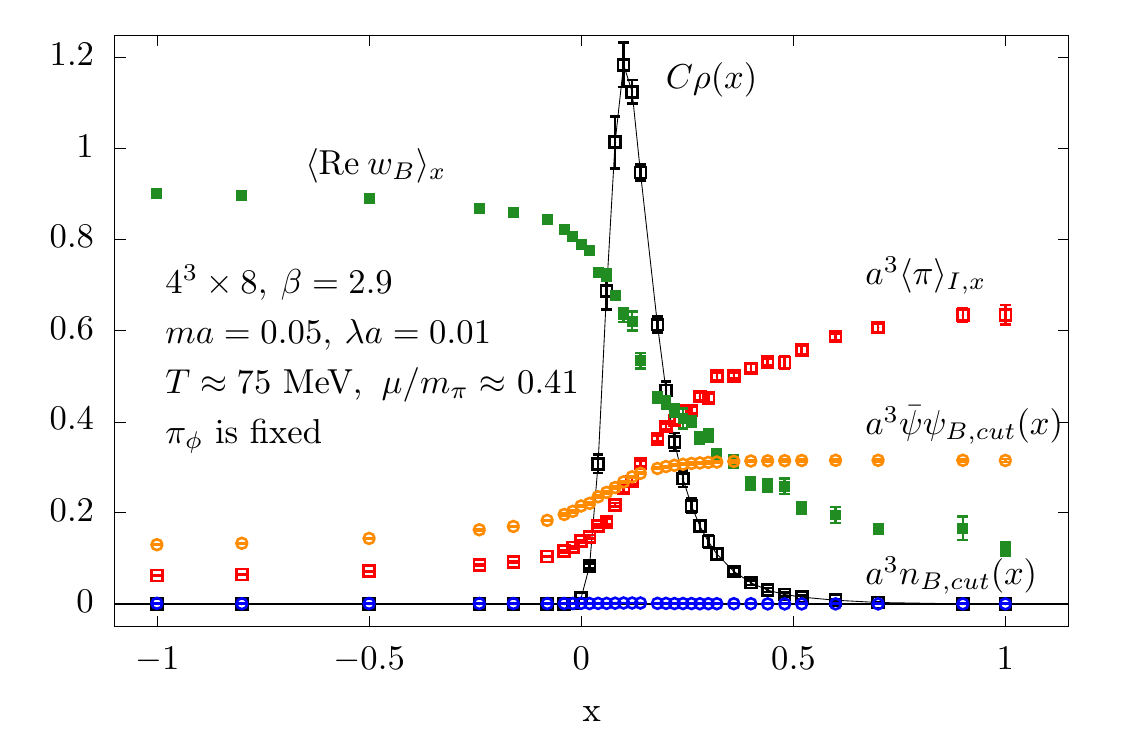}
\includegraphics[scale=0.75]{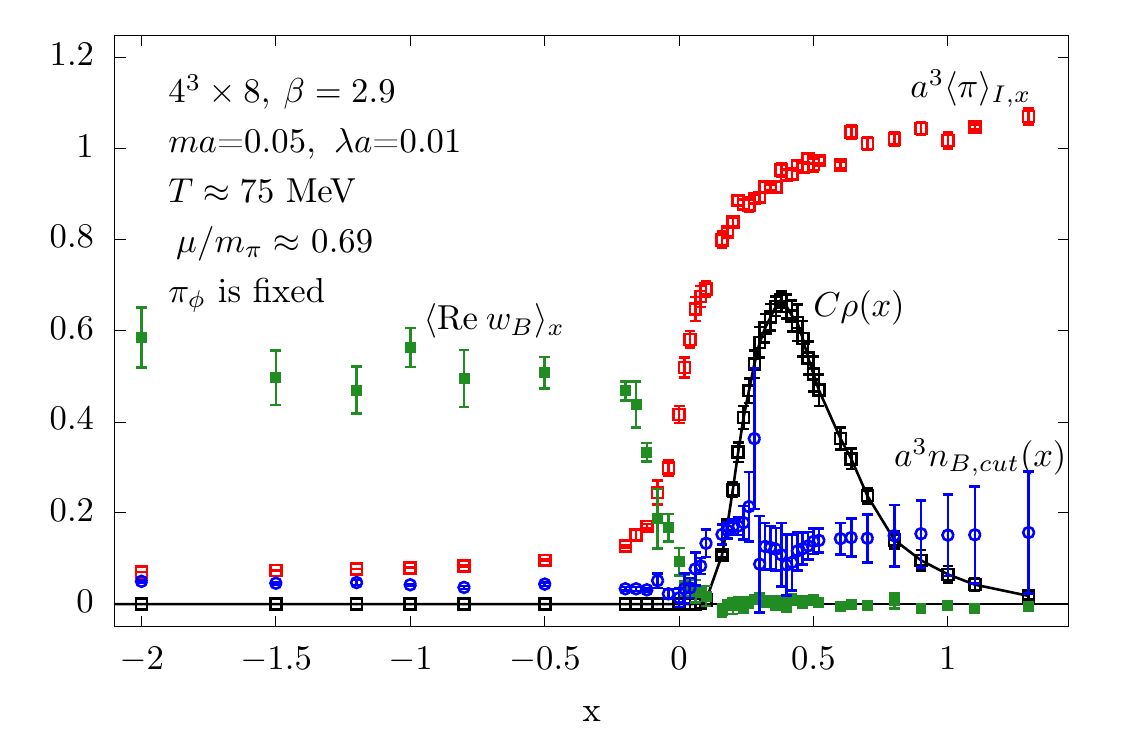}
\caption{
Results obtained by constraining the pion condensate, $\pi_\phi$.
On the horizontal axis, $x$ refers to the fixing value, used in the constraining term.
The black squares show the density of states, $\rho(x)$, multiplied by an irrelevant $C$ constant for visibility.
With red squares, we show pion condensate in the constrained ensembles.
$\langle Re w_B \rangle_x$ refers to the expectation value of the real part of the weights in these ensembles.
$a^3 n_{B,cut}(x)$ is the quark number density calculated with the DoS method, as a function of the upper limit ($x$) of the integrals used in this calculation, defined in Eq. (\ref{cutdef}).
Similarly, $a^3 \bar{\psi} \psi_{B,cut}(x)$ is the (unrenormalized) chiral condensate as a function of the upper limit of the DoS integrals.
}
\label{pion_constr,1}
\end{center}
\end{figure}

After this technical detour, we present results obtained by the method.
In Fig.~\ref{pic_vs_w_pifix}, we show the expectation value of the real part of the weights as a function of the pion condensate for various chemical potentials.
(See also the left panel of Fig.~\ref{phaseavrvspicond_and_voldep} for a similar plot in the case of gauge action fixed simulations.)
As one can see, the decay of the weights as a function of the pion condensate can be well described by an exponential.

In Fig.~\ref{pion_constr,1} we plot various quantities as a function of $x$ for two chemical potential values, below and above $m_\pi/2$.
Note that at the larger chemical potential the peak of $\rho(x)$, the density of states shifts to larger $x$ values, while the average weights are much smaller at larger $x$ values, since the fall-off of the weights is steeper at larger $\mu a$ (see also Fig.~\ref{pic_vs_w_pifix}).

In the DoS integral, the position of the peak of $\langle w_B \rangle_x \rho(x)$ determines which region of $x$ gives the largest contribution to $Z_B$.
The shift of the peak position is determined by the decay of the weights as well as that of $\rho(x)$.
Since the decay of $\rho(x)$ does not depend significantly on $\mu$, based on the $\mu$ dependence of the weights (Fig.~\ref{pic_vs_w_pifix}) it is expected that the shift of the peak is larger as $\mu$ is increased.
This is the motivation to try to include this shift manually by cutting the DoS integrals in Eq.~(\ref{dos_expval}).
In order to analyze the effect of omitting configurations with larger pion condensate, we cut the integrals in the nominator and in the denominator at the same $x$ value, and define $O_{B,cut}(x_c)$ as the ratio of the two cut values, i.e.
\beq
\label{cutdef}
O_{B,cut}(x_c) = \frac{ \int^{x_c}_{-\infty} dx \rho(x) \langle O w_B \rangle_x }{ \int^{x_c}_{-\infty} dx \rho(x) \langle w_B \rangle_x }.
\eeq
The obtained cut results, however, depend on the value where one cuts the DoS integrals.
Since no plateau is visible before the pion condensate starts sharply rising (see Fig. \ref{pion_constr,1}), one could not really select a correct, unique value among the possibilities, the cut results are not unambiguous.

This can be understood by noting that in the range, where $\langle \pi \rangle_{I,x}$ is really small, the value of $\langle O \rangle_{B,cut}(x_c)$ (where $O$ is an arbitrary operator) is predominantly determined by the value of the integrands at $x_c$, as $\rho(x)$ is strongly rising in this region.
In other words, in the range in question, for any observable one gets
\beq
  \frac{ \int^{x_c} dx \rho(x) \langle O w_B \rangle_x }{ \int^{x_c} dx \rho(x) \langle w_B \rangle_x } \approx \frac{ \langle Ow_B \rangle_{x_c} }{ \langle w_B \rangle_{x_c} },
\eeq
which is an estimate for $\langle O \rangle_{B,x_c}$, the expectation value of the $O$ operator in \QCDB\: \textsl{with constraint} characterized by $x_c$.
Of course, $\langle O \rangle_{B,x_c}$ can depend on $x_c$, even though $\langle \pi \rangle_{I,x_c}$ and $\langle \pi_\phi \rangle_{x_c}$ are small.

To overcome the $x_c$ dependence, one could try to carry out another type of reweighting which was introduced in Eq. (\ref{rew3}).
We found that in the range of low $x$ ($x \lesssim -0.2$), where the pion condensate fluctuates less, reweighting with the modified weights of Eq. (\ref{rew3}) including the exponential factor is manageable.
However, this does not eliminate the $x$ dependence.
The results at $\mu < m_\pi /2$ (see the chiral condensate in the upper panel of Fig. \ref{pion_constr,1}) also suggests that in order to have an $x$-independent, correct expectation value, one has to include the configurations with large pion condensate.

In the next section we show that since the gauge action and the pion condensate are slightly correlated, fixing the gauge action also fixes the pion condensate, which has a similar effect on the average weights.
This allows much cheaper simulations (using the gauge action fixing) to be carried out with similar results, we therefore concentrate on those in the following.

\subsection{Fixing the gauge action density}

We now turn to the study of the case when we use the gauge action density as the fixed quantity.
As an illustration, Figure \ref{fig:sg,x} shows the histogram of the gauge action density 
and the expectation value of the real part of the weights as well as the pion condensate 
on the constrained ensemble characterized by $x$, the value at which $s_g$ is constrained.
Fig.~\ref{fig:sg,x} shows that by constraining the gauge action density to smaller values, the pion condensate also becomes small and simultaneously the real part of the weights increases.

\begin{figure}[H]
\begin{center}
\includegraphics[scale=0.69]{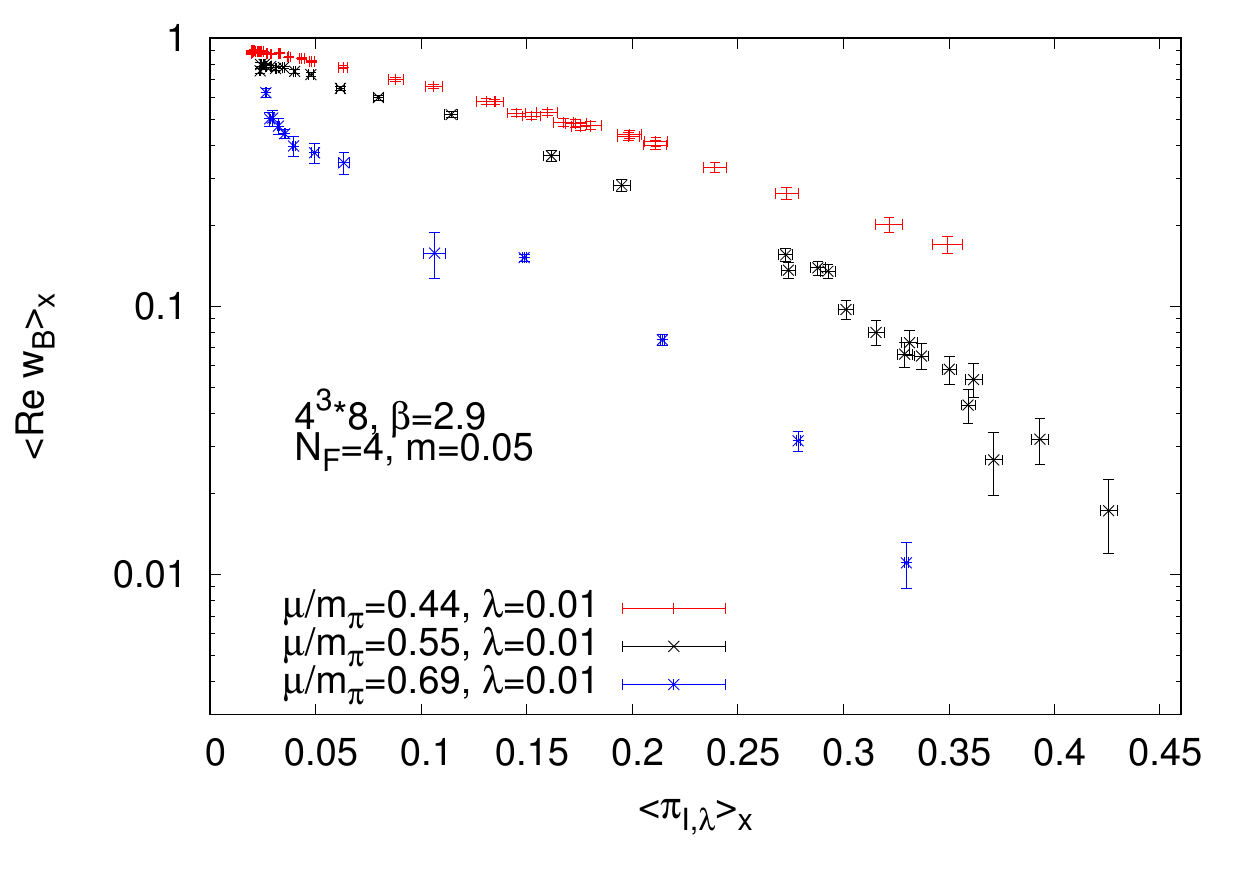}
\includegraphics[scale=0.69]{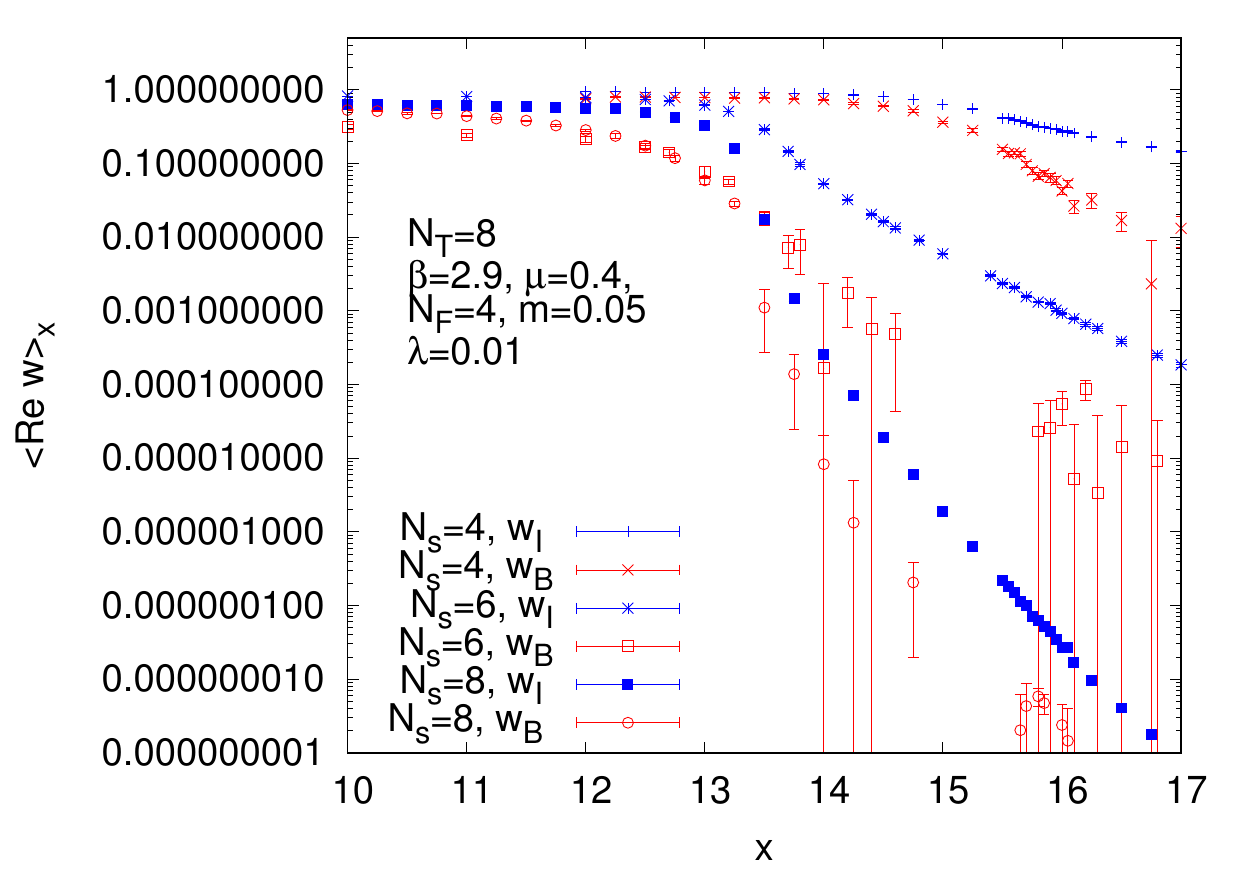}
\caption{
\label{phaseavrvspicond_and_voldep}
Left:
The average weight as a function of the pion condensate in 
simulations with $s_g$ fixing for
several $\mu$ and $\lambda$ values as indicated.
Right:
The average weight as a function of the fixing parameter $x$ 
for several different spatial volumes.
}
\end{center}
\end{figure}

The imaginary part of $\langle w_B \rangle_x$ fluctuates around zero at all $x$.
The correlation between the gauge action density and the pion condensate in \QCDIl\: is also shown in Fig. \ref{fig:sg-pi,corr}.
Although the correlation is weak in the interval of $x$ in which $\rho \sim O(1)$, one can reach configurations with low pion condensate below $x \sim 13.5$.
Similarly to the pion condensate fixing, the real part of the expectation value of the weights as a function of the pion condensate can be described with an exponential with a strongly $\mu$ dependent slope, see the left panel of Fig.~\ref{phaseavrvspicond_and_voldep}.
On the right panel of of Fig.~\ref{phaseavrvspicond_and_voldep} we show the $x$ dependence of the average weights for several volumes.

\begin{figure}[H]
\begin{center}
\includegraphics[scale=0.80]{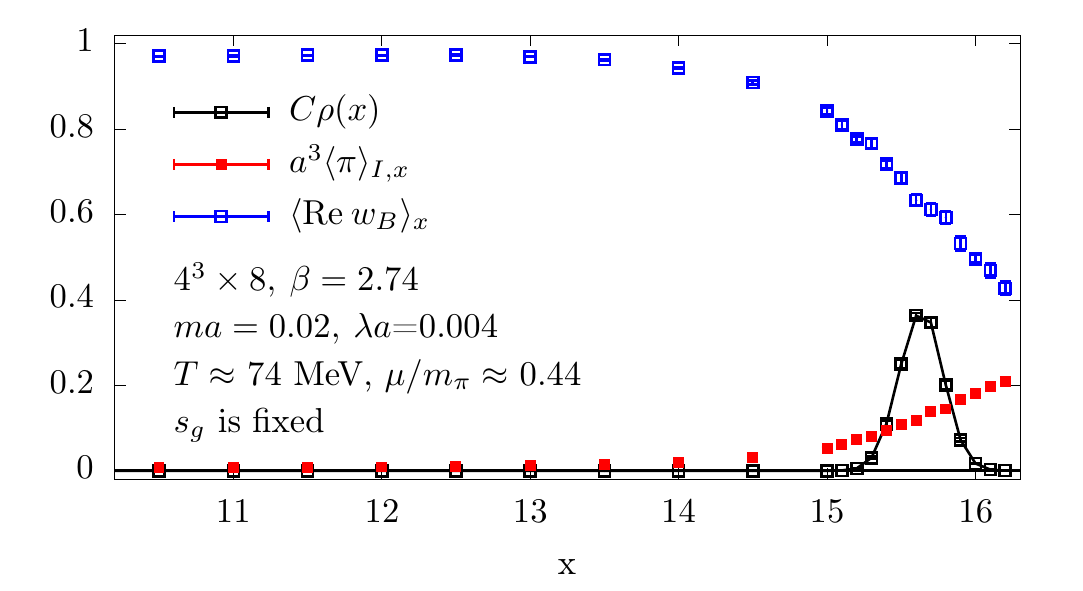}
\includegraphics[scale=0.80]{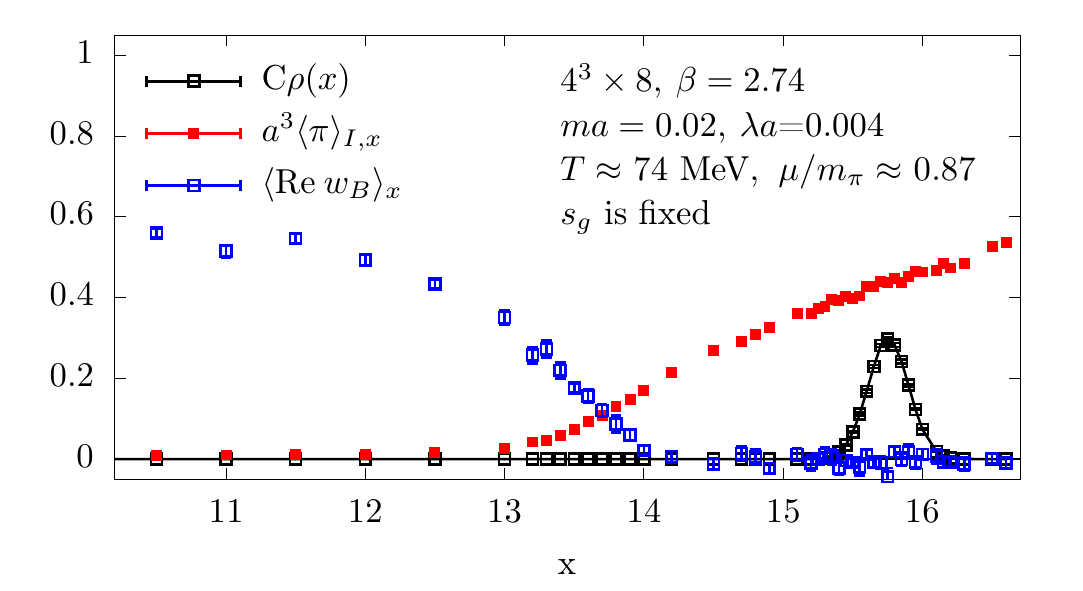}
\caption{
DoS results in the case of $s_g$ fixing at $\mu/m_\pi \approx 0.44$ and at $\approx 0.87$ (left and right panel, respectively).
The real part of the weights goes to zero, as the pion condensate increases.
As $\mu$ is greater, this happens at a lower $x$.
}
\label{fig:sg,x}
\end{center}
\end{figure}

Analyzing the cut DoS integral results (defined in the previous section in Eq. (\ref{cutdef}))
in the case of fixing $s_g$ affirms that when $\mu$ is large ($\mu > m_\pi/2$), the $x$-dependence of the operator dominates the final results for the expectation values 
and one cannot choose a unique cut value, because these depend on $x$.
Moreover, in this case, the cut results could lead to physically problematic results. 
For example the Polyakov loop gets enhanced at low $x$, which suggests that on that configurations, the (approximate) Z(3) symmetry gets broken.
The results for the pion condensate of \QCDBl\: shows in Fig.~\ref{fig:pi_B,cut,8}, as $\lambda$ goes to zero, $\pi_{B,cut}(x)$ also goes to zero at all $x$ even at a larger chemical potential as well -- although with large errors.
This indicates that $\pi_{B,cut}(x)$ is dominated on these lattices by a contribution from the explicit breaking due to finite $\lambda a$.

As in the case of constraining the pion condensate, the findings discussed in the previous paragraph suggest that the configurations with well-behaving weights are not the appropriate configurations to reproduce the expected physics at low temperature.
Thus we abandon the idea of cutting the integrals by hand and in the following, we will analyze the results by calculating the full DoS integrals.

\begin{figure}[H]
\begin{center}
\includegraphics[scale=0.85]{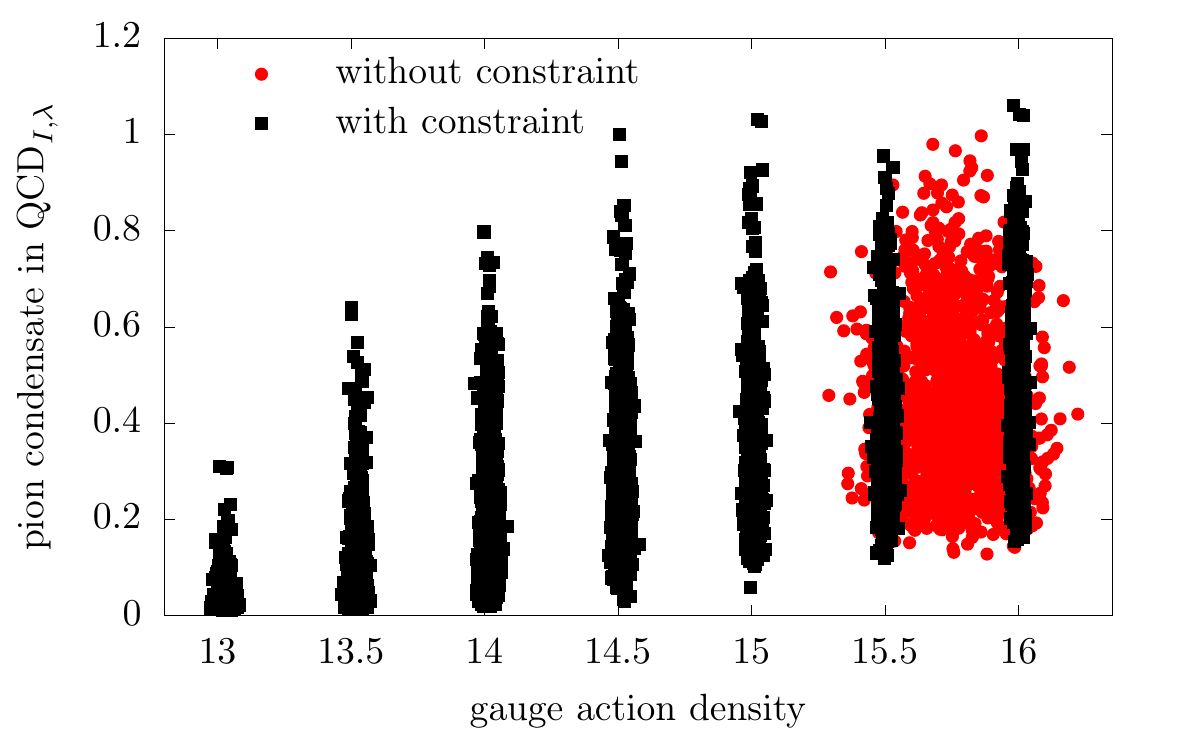}
\caption{
The pion condensate as a function of the gauge action density on individual configurations
at $4^3\times 8$, $\mu/m_\pi \approx 0.87$, $T \approx 74$ MeV, $m_\pi \approx 336$ MeV, $\lambda a= 0.004$.
Shown are the cases, when we constrained the gauge action density to integer and half-integer values (black squares) together with the results obtained from a direct simulation of \QCDIl\,(red circles).
}
\label{fig:sg-pi,corr}
\end{center}
\end{figure}

\begin{figure}[H]
\begin{center}
\hspace*{-0.6cm}
\includegraphics[scale=0.68]{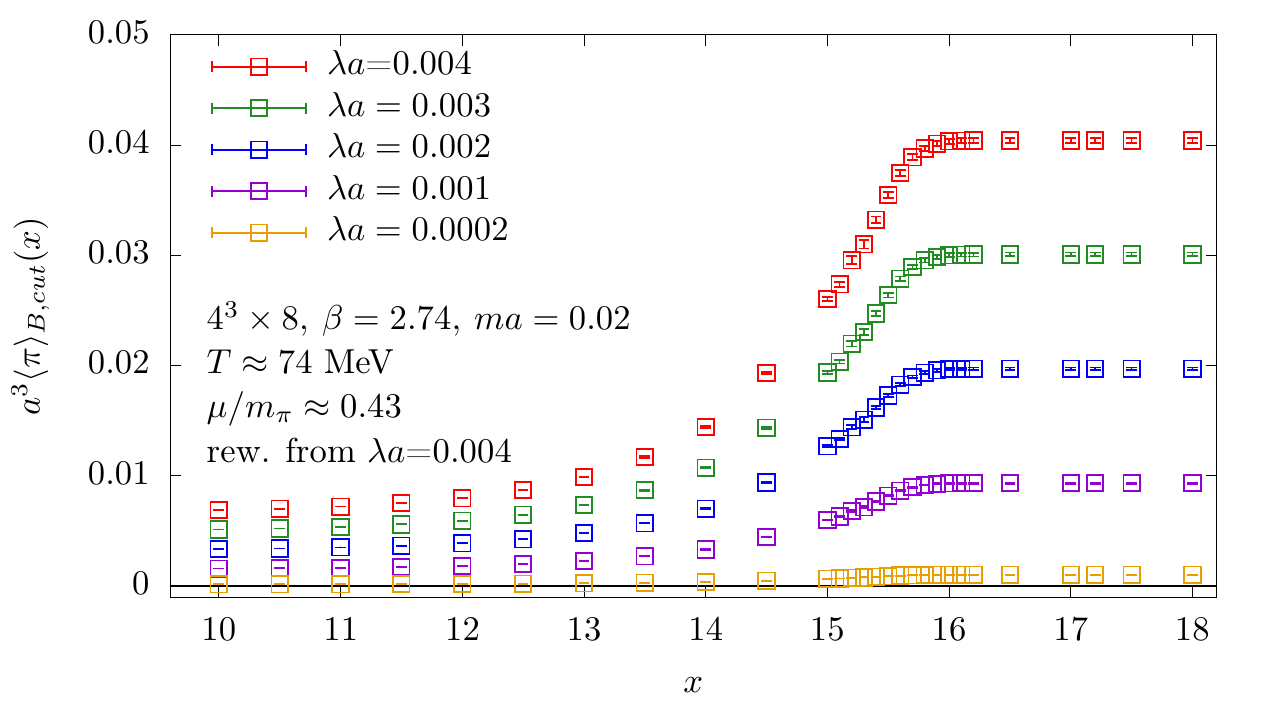}
\hspace*{-0.3cm}
\includegraphics[scale=0.68]{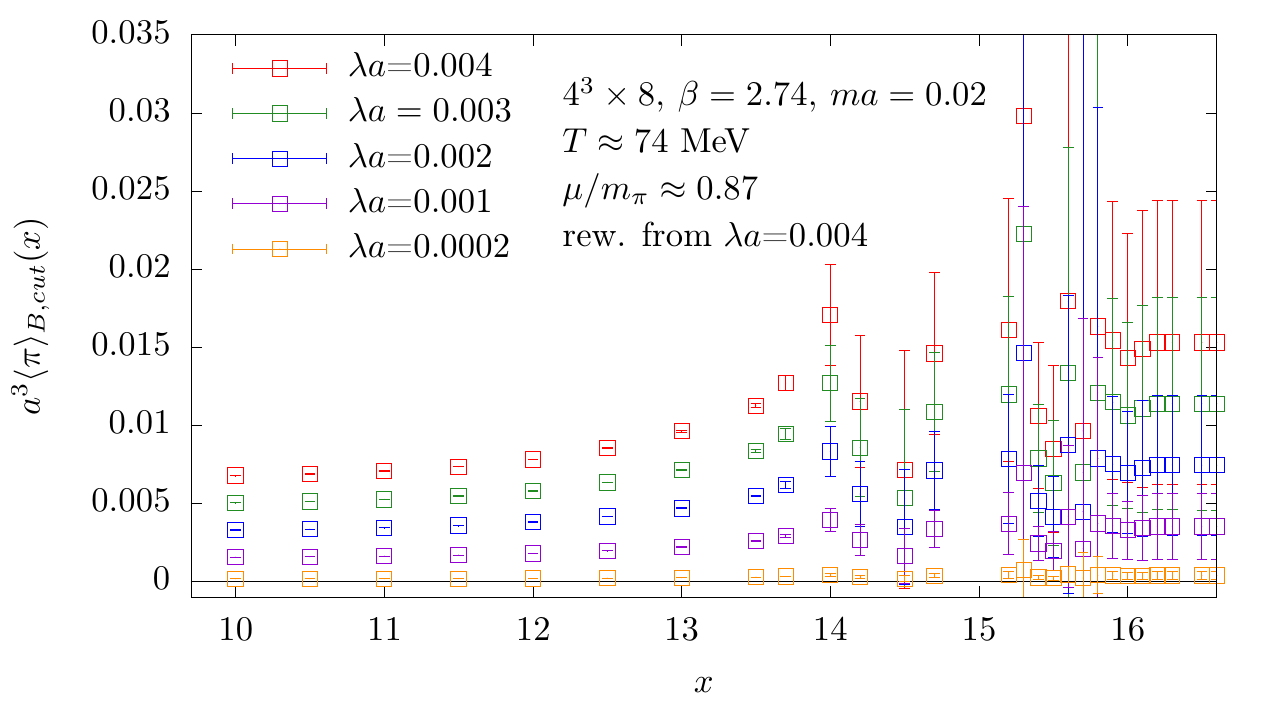}
\caption{
Results for $\langle \pi \rangle_{B, \lambda}(x)$ as a function the upper limit of the DoS integrals in the case of $s_g$ fixing for two different $\mu$ values.
Different colors refer to different 'target' ensembles characterized by $\lambda a$ in the reweighting.
When calculating the full integrals, $\langle \pi \rangle_{B, \lambda}$ is about 2.4 times larger (with huge errors) than the value when cutting the integrals at e.g. 0.
However, both the full integral DoS results and the cut integral results extrapolates to zero in $\lambda a$.
}
\label{fig:pi_B,cut,8}
\end{center}
\end{figure}

As was mentioned above, we demand that the DoS as well as reweighting should provide a positive $Z_B$.
Therefore we try to collect enough configurations to satisfy this criteria at least to a 2 sigma level, which is called our reliability condition.
Since in the case of the DoS, $Z_B=Z_{I,\lambda} \int dx \langle w_B \rangle_x \rho(x)$, while in the case of reweighting, $Z_B=Z_{I,\lambda}\langle w_B \rangle_{I,\lambda}$, we demand $\int dx \langle \rmRe \: w_B \rangle_x \rho(x) > 0$ and $\langle \rmRe\: w_B \rangle_{I,\lambda} > 0$ to hold at 2 sigma, respectively for DoS and for reweighting from \QCDIl.
The positive constant factor $Z_{I,\lambda}$ does not modify the reliablity criteria.
Furthermore, we expect $\int dx \langle \rmIm \: w_B \rangle_x \rho(x) = 0$ and $\langle \rmIm\: w_B \rangle_{I,\lambda} = 0$, to hold, respectively for DoS and for reweighting.

In Figure \ref{full_DoS,Ns4}, we show the results for the quark number density obtained by the DoS method as well as reweighting from \QCDIl\: for the $4^3 \times 8$ ensembles.
Accumulating around $O(10^4)$ configurations at the points where $\rho(x)$ is $O(1)$, we found that the DoS method is reliable up to $\mu a \sim 0.40$ -- and indeed, gives zero quark number density within errors -- on $4^3\times 8$ at pion mass $m_\pi \approx 336$ MeV.
At the finer $4^3 \times 8$ lattice with $\beta=2.9$ and pion mass $m_\pi \approx 437$ MeV, $\mu a \sim 0.45$ can be reached with similar statistics.
These correspond to $\mu / m_\pi \sim$ 0.7 and 0.62, respectively, or $\mu / m_N \sim 0.22 \ldots 0.23$.
Thus, we can reach considerably higher $\mu a$ values than $m_\pi a/2$ at these small lattices.

\hspace{-0.8cm}
\begin{figure}[H]
\begin{center}
\hspace*{-0.8cm}
\includegraphics[scale=0.82]{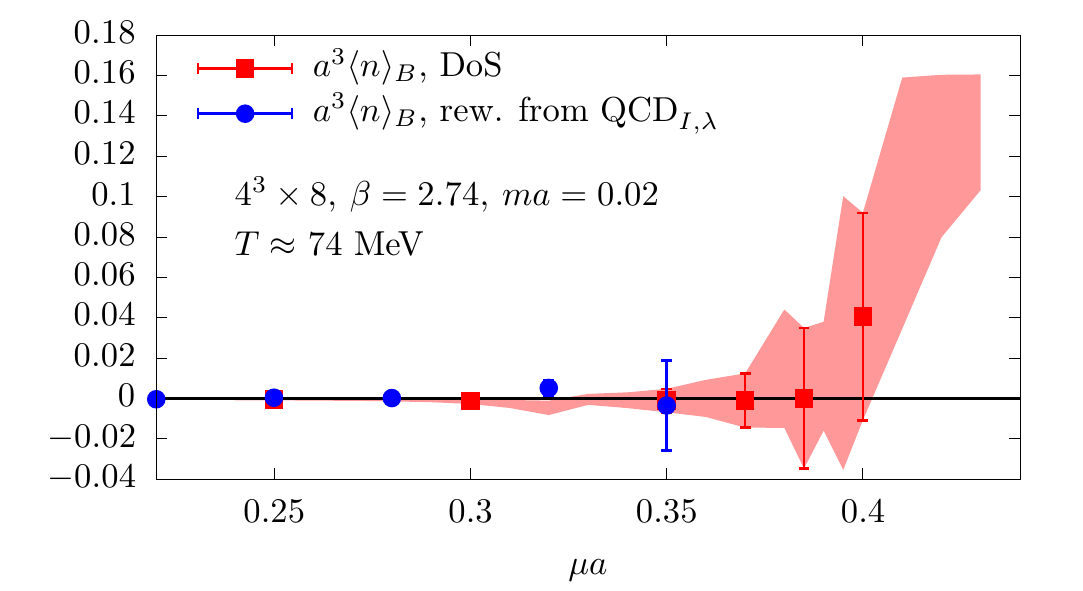}
\hspace*{-0.6cm}
\includegraphics[scale=0.82]{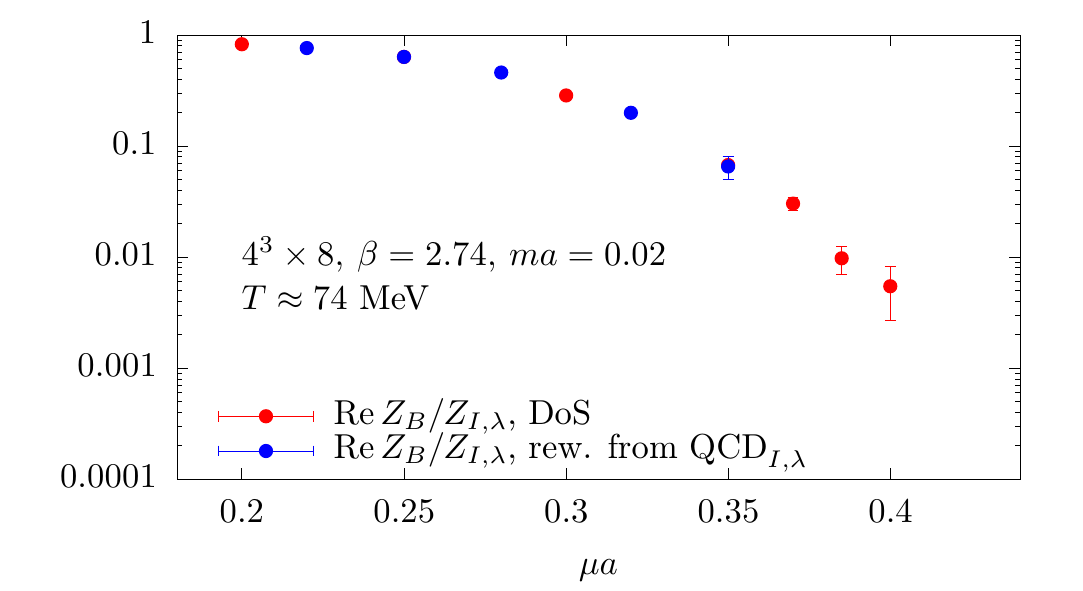}
\caption{
Results obtained by constraining the gauge action density at $4^3 \times 8$, $a m_\pi/2 = 0.2855(5)$.
The band in the left panel is obtained by reweighting from the quark chemical potential $\mu_0$ of \QCDIl\, to $\mu\ne \mu_0$ of \QCDB\, when calculating the DoS integrals.
}
\label{full_DoS,Ns4}
\end{center}
\end{figure}

Two comments are in order.
First, the highest reliably reachable $\mu a$ value certainly depends on the accumulated statistics.
We will elaborate more on this later.
Second, strictly speaking, the pion condensation region of \QCDI\: starts at $m_\pi /2$ only at zero temperature and it can bend toward higher chemical potentials as the temperature increases.
Therefore, to have a reliable comparison, it is important to locate the $a\mu^{(\pi)}_c$ value, where the pion condensation sets in at the given temperature.

\begin{figure}[H]
\begin{center}
\includegraphics[scale=0.73]{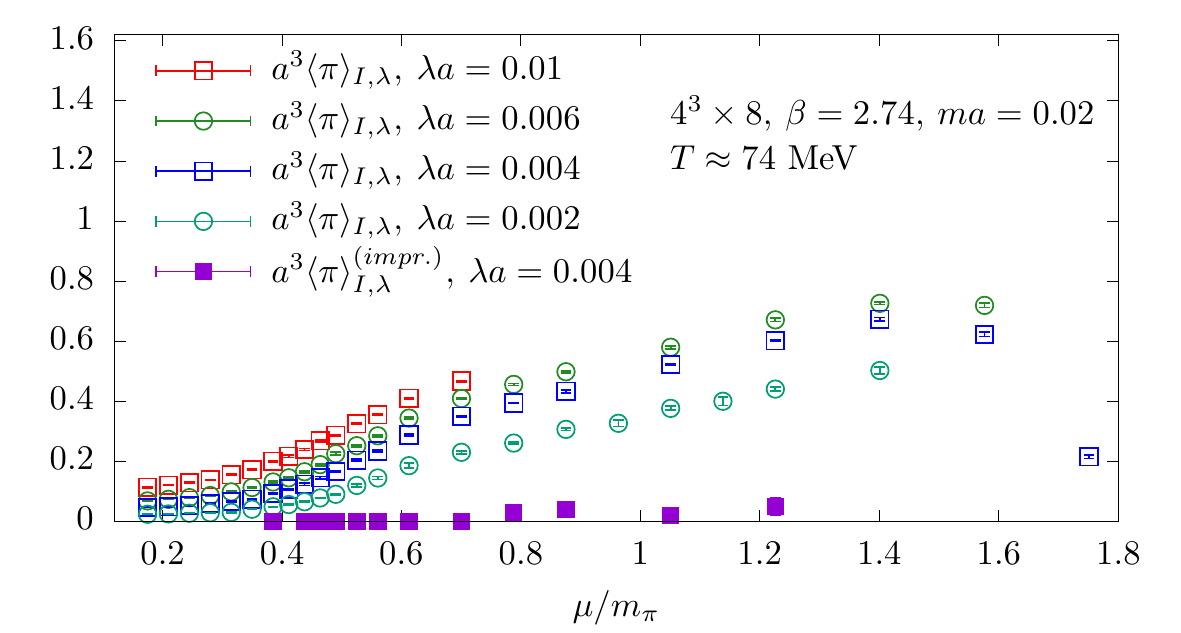}
\includegraphics[scale=0.73]{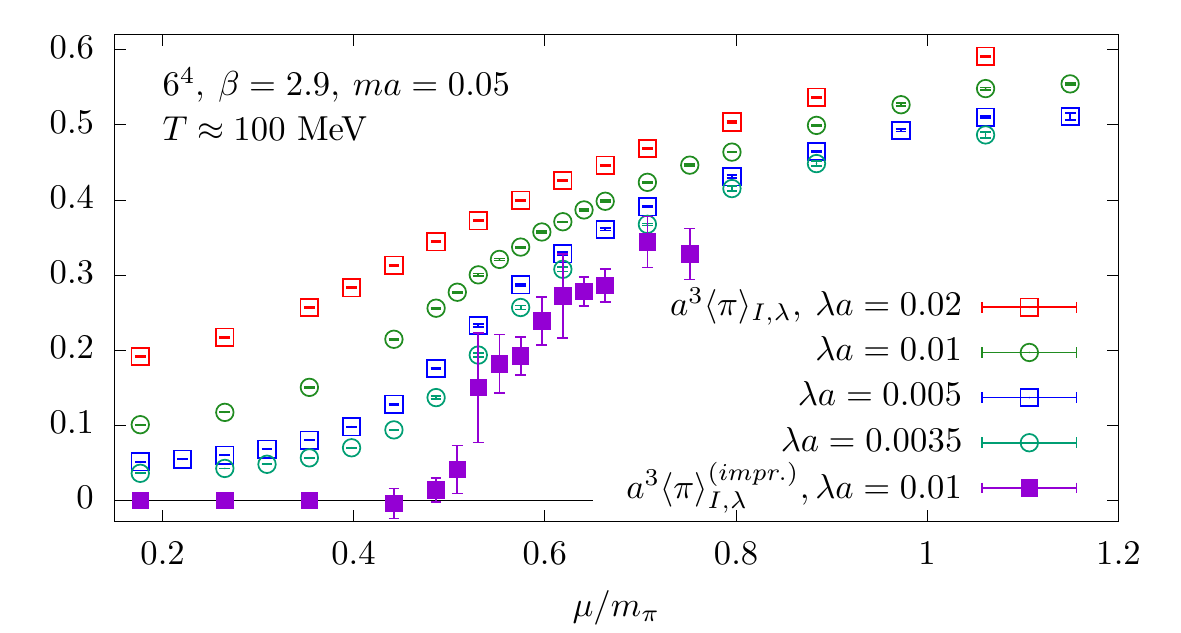}
\caption{
\label{muIc_det}
The pion condensate of \QCDIl\: measured at several $\lambda a$ as a function of $\mu a$ for $4^3 \times 8$ and $6^4$ lattices.
}
\end{center}
\end{figure}

In order to determine this, we carried out simulations at different $\lambda a$ values at several chemical potentials and studied the $\lambda a$ to zero limit.
This extrapolation, however, is not satisfactory to determine precisely $a\mu^{(\pi)}_c$.
Following Ref. \cite{Endrodi:2014lja, Splittorff:2002xn}, we also tried to fit the results by the appropriate formula of chiral perturbation theory \cite{Splittorff:2002xn}, but these fits were rather unreliable, probably due to the fact that the volume is not large enough.
Alternatively, one can obtain $a \mu^{(\pi)}_c(T)$ directly from the lattice simulations with the help of the spectral representation for the pion condensate \cite{Kanazawa:2011tt, Brandt:2016zdy}.
To obtain this, the singular values of the Dirac operator, $\xi_n$, have to be calculated, which are the eigenvalues of $M^\dagger(\mu)M(\mu)$.
Although this approach is valid again if the volume is large enough, following Ref. \cite{Brandt:2016zdy, Brandt:2017zck, Brandt:2017oyy}, one can define $\pi^{(impr.)}$ according to
\begin{align} \label{pi,improved}
  \langle \pi \rangle_{I,\lambda} &= \frac{T}{V}\frac{N_f}{8} 2 \lambda \left\langle \Tr\left(M^\dagger(\mu) M(\mu) + \lambda^2\right)^{-1} \right\rangle_{I,\lambda} = \frac{T}{V} \frac{N_f}{8} 2\lambda \left\langle \sum_n \left( \xi_n^2+\lambda^2 \right)^{-1} \right\rangle_{I,\lambda} \nonumber \\ 
  &\xrightarrow{V\to\infty} \frac{N_f}{4} \left\langle \int d \xi \rho(\xi) \lambda (\xi^2 + \lambda^2)^{-1} \right\rangle_{I,\lambda} \xrightarrow{\lambda \to 0} \frac{N_f \pi}{8} \left\langle \rho(0) \right\rangle_{I,\lambda \to 0} \equiv \langle \pi^{(impr.)}\rangle_{I,\lambda\to 0},
\end{align}
where the spectral density, $\rho(\xi)$ is defined as
\beq
  \rho(\xi) = \lim_{V\to \infty} \frac{T}{V} \sum_n \delta(\xi - \xi_n).
\eeq
In the integral over $\xi$, $\rho(\xi)$ is multiplied by a representation of the Dirac-$\delta$ \mbox{distribution, thus} by taking the $\lambda \to 0$ limit, one arrives at $N_f\pi\rho(0)/8$, which is the improved operator.
Therefore, it is enough to determine the lowest 200-300 singular values, build a histogram for the integrated spectral density,
\beq
  N(\xi) = \int_0^\xi \rho(\xi^\prime) d\xi^\prime,
\eeq
and take the $\xi \to 0$ limit of $a^3 \langle N(\xi) \rangle_{I,\lambda}/\xi$, which after multiplied by $\pi/2$ gives the improved pion condensate.
Unfortunately the approach at $N_s=4$ again cannot be applied probably due to the small volume, but it seems to provide reasonable results at $N_s=6$ (see Fig. \ref{muIc_det}, right panel).
Fig. \ref{muIc_det} justifies that we are in the pion condensed phase on $6^4$, at $\beta=2.9$, $ma=0.05$ above $\mu > m_\pi/2$.

In order to test how the limit of the reliable application of DoS or reweighting changes as we decrease the temperature, we performed simulations at a larger temporal size, namely at $N_t=12$ and $N_t=16$.
We found that although at $4^3 \times 12$, $\beta=2.74$, $ma=0.02$, $\mu a = 0.34 \,\, (\mu/m_\pi \approx 0.6)$, $\rmRe Z_B > 0$ can be satisfied to $\sim$ 9 sigma level by using around 1800 configurations at each $x$, we can not reach even positive $Z_B$ at the 1 sigma level at $\mu a = 0.38 \,\, (\mu/m_\pi \approx 0.67)$ using around 5000-6000 configurations.
Estimating the number of needed configurations using the scaling of absolute errors as the inverse square root of configurations at different $x$ suggests that more than $10^5$ further configurations are needed at each $x$, where $\rho(x) \sim O(1)$, however note that this estimate becomes increasingly unreliable as the relative error of $Z_B$ gets larger, such that a one sigma shift in the average can result in an estimate several orders of magnitude larger.

We accumulated $O(10^3)$ configurations at $4^3 \times 16$, $ma=0.02$, using these, we found that the 2 sigma level condition $\rmRe Z_B > 0$ spoils also in the range $\mu a \sim 0.30\ldots0.36 \,\, (\mu/m_\pi \approx 0.53\ldots 0.63)$.

It is worth to emphasize, that the positiveness of $\rmRe Z_B$ can be satisfied using much less configurations at smaller chemical potential.
In this parameter range the fluctuations of the observable might dominate 
the statistical error of an expectation value (depending on the observable).
This is relevant espically below $\mu=m_\pi/2$, as the phase quenched theory 
also shows a Silver-Blaze behavior and thus reweighting has very little effect. 

\begin{figure}[H]
\begin{center}
\includegraphics[scale=0.69]{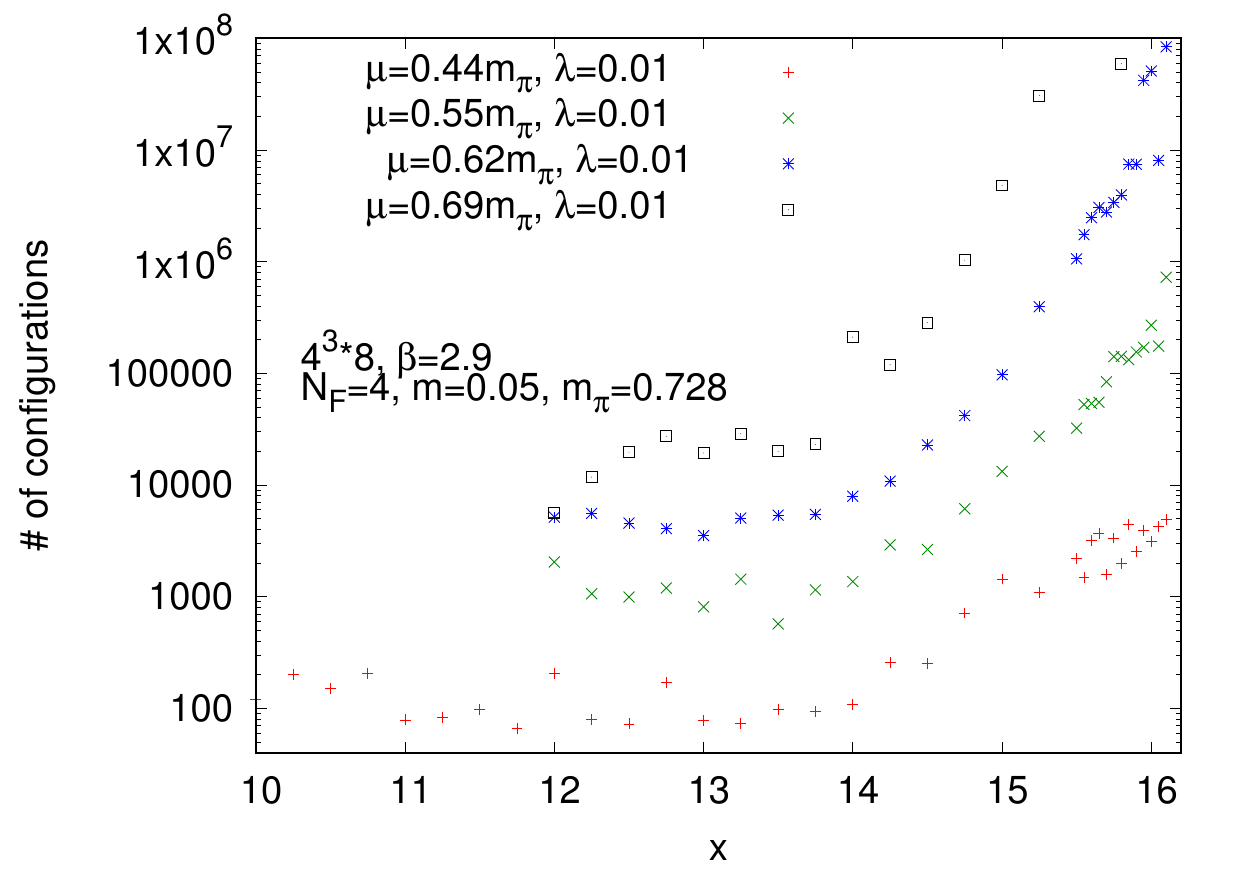}
\caption{
\label{ntopercent}
Number of configurations needed for one percent accuracy measurement 
of $Z_B$, measured on a $4^3 \times 8$ lattice at $\beta=2.9, m=0.05, N_F=4$ at 
various $\mu$ and $\lambda$ values as indicated.
}
\end{center}
\end{figure}

In Fig.~\ref{ntopercent} we show an estimation 
for the number of configurations needed to measure $Z_B$ with the higher 
precision of one percent accuracy on a $4^3\times8$ lattice 
at $\beta=2.9$ and $m=0.05$.  One observes that the number 
of configurations needed increases rapidly with increasing $\mu$. 
In the small $x$ region where the weight average is larger this is approximately constant, but increases at least exponentially with $x$ in the region close to the maximum of $\rho(x)$ (which is in the region $ 15 \le x_{max} \le 16$ for the parameters used in the plot).
A precise calculation of observables at large chemical potentials 
and small temperatures using this method is thus not within the reach 
of current computational capabilities even on very small lattices.

We also estimated the performance of direct reweighting (without introducing a fixing term in the action) for the $4^3 \times 8 $ lattice using $\beta=2.9$ and $m=0.05$.
We observe a similar behavior: we fail to satisfy the $\rmRe Z_B > 0 $ criterion at the 2 sigma level around $\mu/m_\pi \approx 0.76$ even after collecting $ \sim 1.2 \times 10^5$ configurations.
The density is consitent with zero for the chemical potentials where the positiveness of $Z_B$ is satisfied.

Apart from the overlap and the sign problem, we mention a further limitation to safely reach $\mu \sim m_N /3$ on small lattices.
As one might notice at Fig. \ref{muIc_det}, the pion condensate 
starts decreasing around $\mu /m_\pi \sim 1.4 $ (at $\lambda a=0.01$).
This is a saturation effect\cite{Kogut:2002tm}, the isospin density is close to half-filling at these chemical potentials. 
Far from the continuum, $\mu =m_N/3$ might get close to this region where saturation effects dominate the physics, further complicating the issues of reweighting.

\begin{figure}[H]
\begin{center}
\includegraphics[scale=0.65]{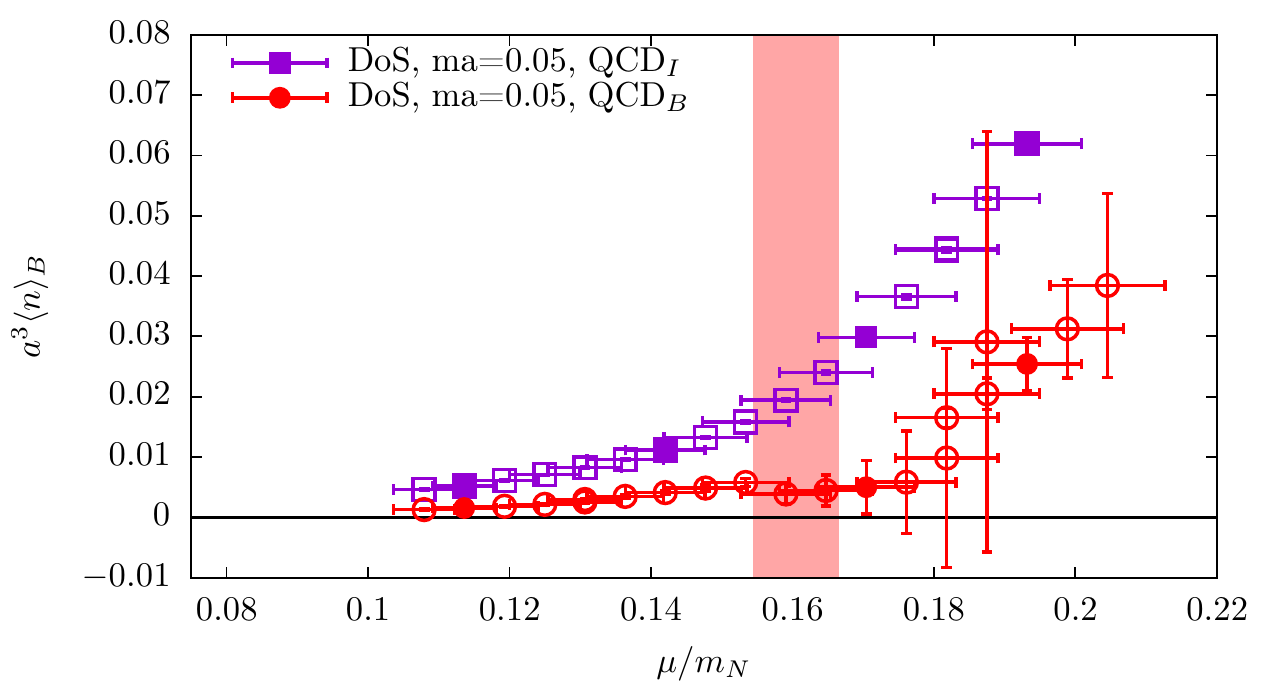}
\includegraphics[scale=0.65]{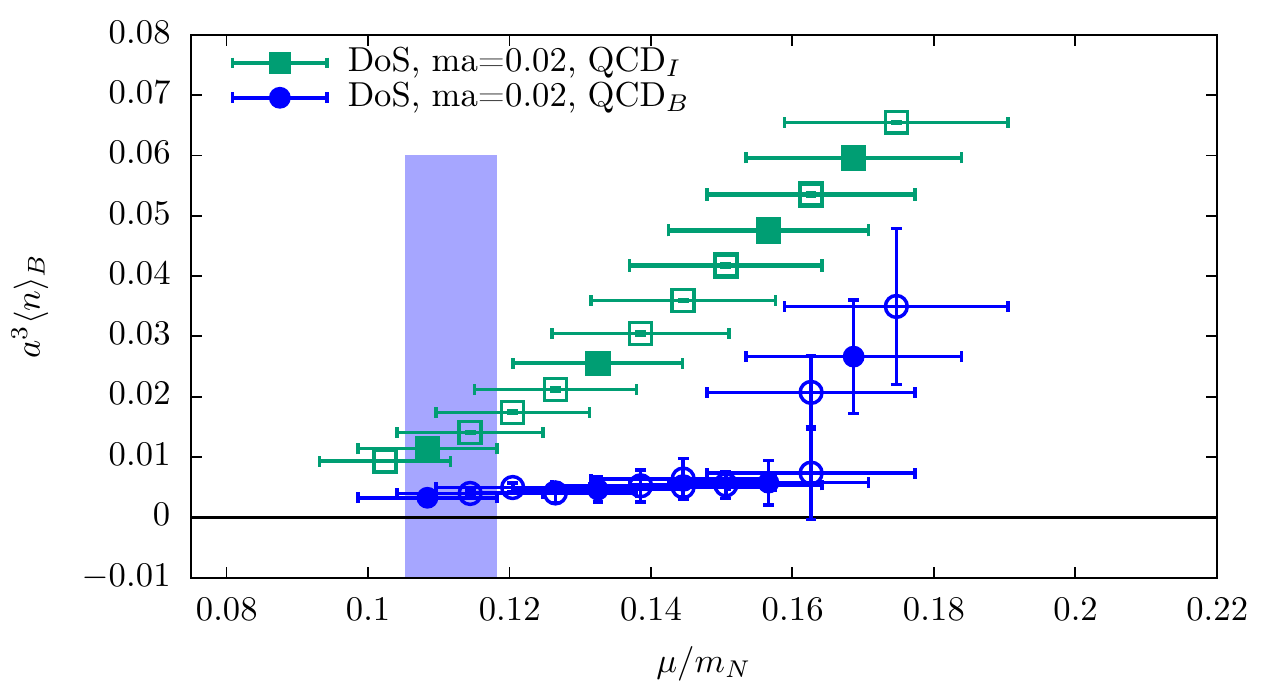}
\vspace{-0.3cm}
\caption{
The quark number density in \QCDB\: and the isospin density in \QCDI\: measured on $6^4$ lattices at $\beta=2.9$, $ma=0.05$ and $m=0.02$.
The huge horizontal errors come from the poor determination of $m_N$.
The open symbols show the results of reweighting from the quark chemical potential $\mu_0$ of \QCDIl\ (where the simulation was carried out), to $\mu\ne \mu_0$, when calculating the DoS integrals.
The red and the blue bands denote the location of $m_\pi/(2m_N)$ for $ma=0.05$ and $ma=0.02$, respectively.
The temperature is around 100 MeV on both ensemble.
}
\label{6x6_dens_onset}
\end{center}
\end{figure}

At $N_s=N_t=6$, $\beta=2.9$, $ma=0.05$, the $\rmRe Z_B > 0$ criteria is valid at more than 2 sigma until the chemical potential range $\mu/m_N \sim 0.2 \ldots 0.22$, see in Fig.~\ref{6x6_dens_onset}.
However, unlike the case of $N_s=4$, it is found that at the last chemical potential 
at which the condition $\rmRe Z_B > 0$ is satisfied more than 2 sigma, i.e. at $\mu a=0.34$, the quark number density, $a^3 \langle n \rangle_B$ becomes nonzero at the $\sim 5$  sigma level.
This chemical potential corresponds to $\mu / m_\pi \sim 0.60$ ($\mu/m_N \sim 0.20$).
At the same lattice size at $T\approx 100$ MeV, but at a smaller quark mass $ma=0.02$, the reliability condition spoils at $\mu/m_N \sim 0.17$ ($\mu/m_\pi \sim 0.76$), so it clearly does not follow a scaling behavior related to $m_\pi$ (Fig. \ref{6x6_dens_onset}).
If that would be the case, based on $(m_\pi a)^2 \propto ma$, one would find the breakdown of the reliability condition around $\mu /m_N \sim 0.14$ (which gives $\mu / m_\pi \sim 0.61$).
For these simulations, at $ma=0.05$, $\mu a=0.34$, we accumulated 4000-5000 configurations at 20 values of $x$, and at $ma=0.02$, $\mu a=0.28$, at around 3000-4000 configurations.
Indications of the onset were found at around $\mu/m_N \sim 0.18$ at $ma=0.05$, and at around $\mu/m_N \sim 0.16$ at $ma=0.02$.

Although to carry out the reweighting at larger chemical potentials would require many orders of magnitude higher statistics, our current results give an indication that the transition line of the nuclear onset is bent to lower chemical potentials at increasing temperatures.

\section{Conclusions}
\label{dos:concl_sec}

In this paper, we have studied the Density of States method and direct reweighting to explore non-zero baryon chemical potential in QCD.
We have included a 'fixing term' in the action of QCD at finite isospin chemical potential, which restricts the values of a chosen operator.
Our investigations are in the cold and dense region of the phase diagram where we sought to observe the Silver-Blaze phenomenon.

In the DoS method, the final results are obtained after calculating the appropriate integrals and normalizing them according to Eq. (\ref{dos_expval}).
When applied at non-zero baryon chemical potential, non-real weights $w_B$ must be included.
In order to classify the results of reweighting as reliable, we have applied the criterion $\rmRe Z_B > 0$ to at least 2 sigma level.

We have tested the fixing of the gauge action, as well as fixing of the pion condensate ($\pi_\phi$ of \QCDIl) and have observed that the results in the two cases behave similarly.
When the pion condensate is constrained to be small, the weight factors $w_B$ are larger.
Constraining the gauge action to a small value lowers the pion condensate.
As a consequence, the weights become larger in this case as well.
Practically it is more economical to use the gauge action fixing as the simulations are much cheaper.

One of the main motivations at the beginning of this work was to investigate whether the DoS integrals receive mainly contributions from configurations that have low pion condensate.
The results revealed that at the parameter range we used, the region of small as well as large pion condensate also contributes to the final results at finite baryon chemical potential.
Although we have indeed observed that the weight factor strongly depends on the pion condensate, the shift in the peak of $\rho(x) \langle w_B \rangle_x$ is moderate on small lattices.
At larger volumes, the weights decay faster, but $\rho(x)$ also becomes narrower, which results in a negligible shift of the peak position.
We have investigated whether one can improve the situation by cutting the integrals over $x$ manually, and only allow configurations with small pion-condensate to contribute, as suggested in \cite{Aoki:2014mta}.
Although this way well-behaving weights can be obtained even at $\mu > m_\pi/2$, there is no such region of the upper limit of the integrals in which the cut observables are constant.
Therefore, we conclude that cutting the DoS integrals is not a viable procedure to determine the expectation values of the studied observables.

The sign problem becomes severe around $\mu \approx m_\pi/2$.
We have estimated the number of configurations one needs in order to overcome the sign problem slightly over this value.
This number grows very quickly with the chemical potential, such that one can not go deep into the $\mu>m_\pi/2$ region even on very small lattices.
However, at $6^4,\: \beta=2.9,\: T \approx 100$ MeV, we have managed to apply the DoS method with reweighting classified to be reliable, and found indications of the baryonic onset at this temperature.
This observation would imply that at higher temperatures the baryonic onset happens at lower chemical potentials than the zero temperature critical chemical potential $\mu \approx m_N/3$. 

Finally, we note that both the DoS and direct reweighting from \QCDIl\: provide results consistent with our expectations as long as they are classified as reliable.

\acknowledgements

This research was funded by the DFG grants Heisenberg Programme (SE 2466/1-1), Emmy Noether Programme (EN 1064/2-1), SFB/TR55; by the New National Excellence Program of the Ministry of Human Capacities of Hungary; and was partially supported by the Hungarian National Research, Development and Innovation Office — NKFIH grants KKP126769 and K113034.
The authors gratefully acknowledge the Gauss Centre for Supercomputing e.V. (www.gauss-centre.eu) for funding this project by providing computing time on the GCS Supercomputers JUQUEEN \cite{juqueen} and JUWELS at Jülich Supercomputing Centre (JSC).

\bibliographystyle{unsrt}
\bibliography{mybib}

\end{document}